\documentclass[aps,prab,reprint,preprintnumbers,amsmath,amssymb,floatfix,longbibliography]{revtex4-2}
\usepackage{graphicx,booktabs,bm}
\usepackage[colorlinks=true,allcolors=blue]{hyperref}
\graphicspath{{figures/}}
\newcommand{\pcell}[2]{\parbox[t]{#1}{\raggedright #2}}
\begin{document}
\title{Slipstream: nonlinear beam dynamics in PyTorch}
\author{J. K. Jones}
\author{A. D. Brynes}
\author{M. King}
\author{H. L. Owen}
\author{A. Pollard}
\affiliation{
    Accelerator Science \& Technology Centre, STFC Daresbury Laboratory,    Warrington, United Kingdom and the Cockcroft Institute,
    Warrington, United Kingdom
}\date{November 24th, 2026}
\begin{abstract}
\textsc{Slipstream} is a \textsc{PyTorch} package for nonlinear beam dynamics in storage rings, linacs and photoinjectors. It provides CPU and GPU particle tracking, field-map injector models with space charge, and automatic differentiation along supported tracking paths. Comparisons with \textsc{elegant}, \textsc{ASTRA}, \textsc{GPT} and other codes assess accuracy and computational cost for nonlinear ring tracking, optics matching and photoinjectors. Ring tunes agree with \textsc{elegant} to $10^{-9}$. For the tested ring workloads with tens of thousands of particles, GPU tracking in float32 is hundreds of times faster than the single-threaded CPU implementations compared here. Automatic differentiation reduces the time for 24-parameter optics matching by an order of magnitude compared to finite differences. Using a high-brightness photo-injector as an example, final horizontal emittances agree with \textsc{ASTRA} to about 1\% at 20 and 100~pC and differ by $\approx$4\% at 250~pC. For $2^{18}$ macroparticles, \textsc{Slipstream}'s faster default settings take 14~s on an RTX 3050 GPU and give emittances within 2\% of that of \textsc{ASTRA}. Additional tests examine coherent synchrotron radiation, dynamic aperture and resistive-wall impedance. The package was developed predominantly by AI coding assistants under human direction. These comparisons identify where its models, GPU execution and derivatives are useful for repeated simulations in scans, fits and optimisation.
\end{abstract}

\maketitle

\section{Introduction}
\label{sec:intro}

Accelerator design, commissioning and optimisation often require beam-dynamics calculations to be repeated over many machine settings. Digital-twin applications place similar demands on simulation~\cite{brynes2026,king2025,edelen2016,edelen2024}. The calculation must be fast enough for a scan or fit while retaining the physical effects that determine the observable of interest. A linear transport model can meet this need for some optics tasks, whilst non-linear ring dynamics and space-charge-dominated injectors require more detailed models.

Existing codes combine these capabilities in different ways, and GPU execution is available in several of them. \textsc{Cheetah} is a \textsc{PyTorch} code for differentiable beamline transport~\cite{kaiser2024}; its linear models can be reduced to efficient transfer maps, and recent releases add the nonlinear element options used in the comparisons below, as well as space-charge kicks~\cite{dhamrait2026}. Here, we use differentiable to mean that the tracking is implemented with operations supporting automatic differentiation (AD), so that exact gradients of an output with respect to all inputs can be obtained by back-propagation through the simulation, at a cost that does not grow with the number of inputs (it grows with the number of outputs) and is small compared with the many forward runs needed by finite differences (FD). \textsc{SciBmad} and \textsc{JuTrack} (both \textsc{Julia}) and the linac framework \textsc{HELIX} (\textsc{PyTorch}, developed for the PIP-II hadron linac, with a particle-in-cell space-charge solver) also provide differentiable tracking, using different element models and space-charge treatments~\cite{signorelli2026,wan2025,qiang2023,pathak2026}. \textsc{Xsuite} and \textsc{ImpactX} are established tracking codes with nonlinear maps and collective effects, designed for both CPU and GPU execution~\cite{iadarola2023,huebl2022,mitchell2024}. Because these choices are implemented differently, the same nominal lattice need not represent the same numerical map in every code, which matters when comparing both accuracy and speed.

A photoinjector provides a demanding example. The emittance delivered to a free-electron laser is largely determined in the gun and first accelerating structure, where emission, space charge, radio-frequency (RF) fields and solenoid focusing interact~\cite{kim1989,carlsten1989,serafini1997,bazarov2009,dowell2009}. \textsc{ASTRA} and \textsc{General Particle Tracer} (\textsc{GPT}) are established tools for these calculations~\cite{floettmann2017,vangeer2001}. Repeated injector simulations are useful for finding working points, assessing jitter and fitting machine parameters.

\textsc{Slipstream} combines nonlinear ring and linac transport with field-map photoinjector tracking in \textsc{PyTorch}. Injector simulation includes quasi-static space charge and a cathode image field; the ring and linac trackers include magnetic transport and selected collective-effect models. The package reads \textsc{elegant}~\cite{borland2000} and \textsc{ASTRA} input files and supports CPU and CUDA GPU execution. Supported physics calculations also retain their dependence on machine parameters, allowing automatic differentiation to supply derivatives for fitting and optimisation, though currently emission timing and aperture losses remain outside the differentiable path.

We ask three practical questions: how closely the models reproduce established numerical results, how execution time changes with particle number, and when derivatives reduce the time required for optimisation. Ring tracking and optics matching test the transport and derivative paths. A space-charge-dominated drift and the CLARA gun~\cite{angalkalinin2020} and first accelerating structure test the collective models and the numerical settings needed for convergence. Further comparisons examine coherent synchrotron radiation (CSR), booster dynamics, intrabeam scattering and resistive-wall impedance.

The development also provides a case study of AI-assisted accelerator simulation. Coding assistants wrote most of the implementation and accompanying software under the authors' direction, covering physical models, input formats and comparison workflows. The paper examines what this process produced and how its numerical results can be checked independently. Section~\ref{sec:ai-development} describes the workflow and concrete examples. Sections~\ref{sec:methods} and~\ref{sec:setup} describe the models and benchmark protocols, followed by the main results in Sec.~\ref{sec:results}, additional comparisons in Sec.~\ref{sec:further}, and their practical implications in Sec.~\ref{sec:discussion}.

\section{Physical models and numerical methods}
\label{sec:methods}

\begin{figure*}[t]
\includegraphics[width=\textwidth]{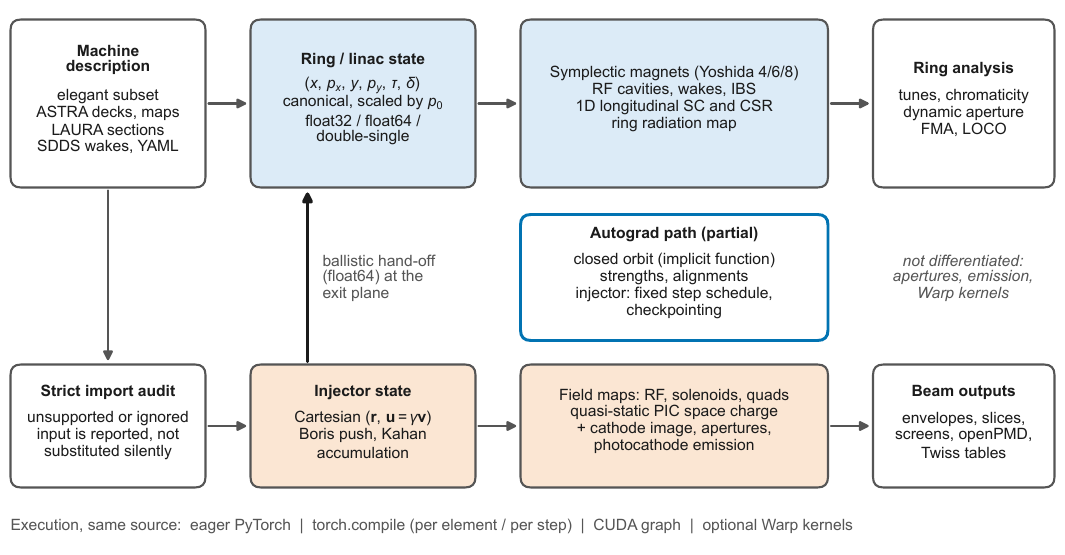}
\caption{Tracking paths and state descriptions in \textsc{Slipstream}. Blue boxes show ring and linac transport; orange boxes show Cartesian injector tracking. Automatic differentiation covers the supported ring maps and closed-orbit calculation, and the injector calculation with a fixed step schedule. Emission timing and aperture losses are not differentiated. In the analysis box, FMA denotes frequency-map analysis and LOCO denotes linear optics from closed orbits.}
\label{fig:workflow}
\end{figure*}

\textsc{Slipstream} uses two descriptions of the particle bunch (Fig.~\ref{fig:workflow}). Ring and linac trackers advance a tensor of six-dimensional states through lattice elements, applying each element to all particles as a batched operation. The injector instead advances Cartesian positions and proper velocities through applied and collective fields. An explicit coordinate conversion joins the two trackers. \textsc{PyTorch} provides the main implementation~\cite{paszke2019}; optional \textsc{Warp} kernels accelerate selected calculations~\cite{macklin2022}.

\subsection{State vector and conversions}
\label{sec:state}

Ring and linac states are $(x,p_x,y,p_y,\tau,\delta)$. The momenta $p_x$, $p_y$ and $p_z$ are normalised to the reference momentum $p_0$, $\tau$ is the arrival delay expressed as a length in metres and is positive for late arrival, and $\delta=(p-p_0)/p_0$. The longitudinal momentum is
\begin{equation}
 p_z=\sqrt{(1+\delta)^2-p_x^2-p_y^2},
 \label{eq:pz}
\end{equation}
and the slope is $x'=p_x/p_z$, with the corresponding expression for $y'$. At finite energy, $\delta$ differs from the energy variable conjugate to arrival time.

The injector exit bunch is converted from a common time to a common longitudinal observation plane by ballistic propagation. This common-$s$ conversion is evaluated in float64 before casting to the working precision. In particular, forming $u/u_0-1$, where $u$ and $u_0$ are the particle and reference proper-speed magnitudes, in float32 can lose significant digits when the two values are nearly equal.

\subsection{Magnetic transport}
\label{sec:magnetic}

In a straight drift, $x$ changes by $\Delta s\,p_x/p_z$, and similarly for $y$. The arrival coordinate changes by $\ell/\beta-\Delta s/\beta_0$, where $\ell$ is the particle path length and $\beta c$ and $\beta_0c$ are the particle and reference speeds. A sector bend of curvature $h$ uses a closed-form drift in canonical variables. Its trigonometric expressions are rearranged to avoid subtracting nearly equal quantities when $h\Delta s$ is small, which is important in float32.

Multipole kicks are
\begin{equation}
 \begin{aligned}
 F_y &= h + k_1 x + \tfrac12 k_2 (x^2-y^2) + \dots,\\
 F_x &= k_1 y + k_2 xy + \dots,\\
 \Delta p_x &= -\Delta s\,(1+hx)F_y,\\
 \Delta p_y &= \Delta s\,(1+hx)F_x,
 \end{aligned}
 \label{eq:kicks}
\end{equation}
where $k_1$ and $k_2$ are the normalised quadrupole and sextupole strengths. Higher multipole orders follow the same expansion. Symmetric drift--kick slices compose each element, with separate slice counts for quadrupoles and sextupoles. For an order-$n$ symmetric map $S_n$, Yoshida's triple-jump construction~\cite{yoshida1990},
\begin{equation}
 \begin{aligned}
 S_{n+2}(\Delta s)&=S_n(x_1\Delta s)\,S_n(x_0\Delta s)\,S_n(x_1\Delta s),\\
 x_1&=\frac{1}{2-2^{1/(n+1)}},\quad x_0=1-2x_1,
 \end{aligned}
 \label{eq:yoshida}
\end{equation}
raises the integration order. Orders 4, 6 and 8 are available. Edge maps include the gap, fringe-field integral and Brown's second-order terms~\cite{brown1982}, following \textsc{elegant}'s implementation~\cite{borland2000}.

An exact-dipole option incorporates the dipole field into the drift map to improve float32 accuracy. In the booster comparison (Sec.~\ref{subsec:booster}), the float32 dynamic-aperture area then agrees with \textsc{elegant} to 3\%. At the default of 8 slices per bend, the discretised combined-function bend can displace the nominal reference orbit. The tracker subtracts the displacement produced by each ideal element, analogous to \textsc{elegant}~\cite{borland2000}.

\subsection{RF and collective effects in rings and linacs}
\label{sec:rf}

Ring RF cavities can act once per turn as a lumped kick or at their individual lattice positions. Travelling-wave structures have a fixed phase velocity, allowing a low-energy particle to slip in phase. The linac tracker recentres the reference momentum on the bunch centroid after each element, following \textsc{elegant}. A separate mode uses a fixed design-energy profile for jitter studies.

The linac includes a local, long-wavelength model of longitudinal space charge (LSC)~\cite{saldin2004}. Its one-dimensional CSR model includes a steady-state wake and an optional entrance transient, evaluated at the centre of each dipole slice~\cite{borland2001,saldin1997}. The wake behind the dipole exit, which carries the exit transient, is taken from the closed-form drift model of Stupakov and Emma~\cite{stupakov2002} as in \textsc{elegant}'s \texttt{CSRDRIFT}, from a kernel tabulated from the retarded field of a line charge, or, more simply, from a steady-state kick truncated at the overtaking length. Convolution with the line density gives the wake potential. The steady-state calculation uses either pairwise quadrature of the $|\Delta s|^{-1/3}$ wake kernel, with cost $O(N^2)$ for $N$ particles and a softening length $\epsilon$ that removes the singularity at zero separation, $|\Delta s|^{-1/3}\to(\Delta s^{2}+\epsilon^{2})^{-1/6}$, or a fixed histogram grid, with cost $O(N+B\log B)$ for $B$ bins. The transient and drift kernels use the fixed grid. Section~\ref{sec:csr} compares these approximations.

Ring synchrotron radiation acts through a once-per-turn damping and excitation map. Intrabeam-scattering (IBS) rates follow Bane and Bjorken--Mtingwa~\cite{bane2002,bjorken1983}. Short-range wakes are applied as per-cell wake-potential tables $W(t)$ (V/C longitudinal, V/C/m dipole), read from \textsc{SDDS} files or generated from a tabulated impedance by Gaussian-probe inverse FFT. Bunch charge is binned onto the table's time grid, convolved with the wake by FFT, and the resulting voltage is read back per particle.

\subsection{Injector: applied fields and space charge}
\label{sec:injector}

\paragraph*{Particle advance.} The injector advances Cartesian positions and proper velocities $\bm u=\gamma\bm v$ with the relativistic Boris push scheme~\cite{qin2013}. A pure magnetic rotation (e.g. a solenoid) preserves $|\bm u|$, allowing $\gamma$ to be evaluated once per step. At gun exit energies, a longitudinal space-charge increment $a=(q/m)E\Delta t$ can be smaller than the float32 spacing between stored values of $u$. Direct addition would then discard the increment. Positions and momenta therefore use Kahan compensated summation~\cite{kahan1965},
\begin{equation}
 y=a-c,\quad t=u+y,\quad c\leftarrow(t-u)-y,\quad u\leftarrow t,
 \label{eq:kahan}
\end{equation}
which carries the rounding correction $c$ between steps.

\paragraph*{Applied fields.} On-axis maps describe standing-wave cavities, travelling-wave structures stored as one repeating period, solenoids including bucked combined maps~\cite{scott2017}, quadrupoles, dipoles with Fermi-profile fringes, and correctors. A paraxial radial expansion evaluates fields away from the axis. For example, a solenoid uses $B_r=-\tfrac r2\,\mathrm{d}B_0/\mathrm{d}z$ and $B_z=B_0-\tfrac{r^2}{4}\,\mathrm{d}^2B_0/\mathrm{d}z^2$, where $B_0(z)$ is the on-axis field. Automatic phasing finds each cavity's maximum on-axis energy gain by a phase scan with parabolic refinement.

\paragraph*{Space-charge solve.} Cloud-in-cell weights deposit charge on a grid in the bunch rest frame, stretching the longitudinal coordinate by the mean Lorentz factor $\gamma$. The solver calculates the electrostatic potential from $\nabla^2\phi'=-\rho'/\epsilon_0$ and interpolates its field to the particles. The transverse lab-frame force is
\begin{equation}
 \bm F_\perp=\frac{q}{\gamma}\,\bm E'_\perp.
 \label{eq:boost}
\end{equation}
This quasi-static transformation includes the electric-field boost and the compensating magnetic force~\cite{vay2008,qiang2006}.

The default particle-in-cell (PIC) solver assumes axial symmetry. It uses conservative radial finite differences with a grounded wall at the radial edge of the box, and a zero-padded (Hockney) FFT in the longitudinal direction to give an open longitudinal boundary~\cite{birdsall2004,hockney1988}. The alternative Cartesian solver uses a cell-averaged integrated Green's function and open-boundary FFT convolution~\cite{qiang2006}. Both represent the conducting cathode by depositing an image bunch of opposite charge through the same solver while the bunch is near the cathode. The cylindrical box extends radially to the outermost particle plus $6\sigma_r$ and longitudinally (in the rest frame) to the bunch extent plus $6\sigma_z$, with a further 25\% longitudinal slack; the Cartesian box adds 5\% of the particle span on each side.

\paragraph*{Emission.} Each macroparticle becomes active at the first step at or after its assigned emission time. An optional photocathode model calculates emitted charge from the laser pulse and extraction field using the three-step model of Dowell and Schmerge~\cite{dowell2009}. The finite step quantises the emission time; this is important for both convergence and derivative coverage.

\paragraph*{Structure wakes.} Tabulated monopole wakes enter as localised longitudinal kicks, using the model of \textsc{ASTRA}'s \texttt{\&WAKE} tables~\cite{floettmann2017,dohlus2012}. When the bunch centroid reaches a kick, the particle energy changes by $-qV(z)$, where $V(z)=\sum_j q_jw(z_j-z)$ sums the contributions of particles ahead. The wake function $w$ is tabulated in V/C. The potential is evaluated on uniform bins with linear charge weights. A structure with $n$ cells receives $n$ kicks when its table describes one cell. Transverse structure wakes are not yet modelled in the injector.

\paragraph*{Apertures and diagnostics.} Apertures are tested along the straight path swept by each particle during a step. Screens record the first forward crossing by interpolation. Observed and lost charges are reported separately. 

\subsection{Differentiation and batching}
\label{sec:diff}

Magnet strengths, corrector kicks, offsets and rolls can remain tensors in the ring maps. Tensors with shape $(M,1)$ allow $M$ machine settings to be tracked in one batch. Here a machine parameter, also called a knob, is an input varied in a scan, sensitivity calculation or optimisation.

The closed orbit satisfies $\bm r(\bm z_{\rm co};\bm\theta)=\mathcal M(\bm z_{\rm co};\bm\theta)-\bm z_{\rm co}=0$, where $\mathcal M$ is the one-turn map and $\bm\theta$ the parameter vector. The Newton solver retains the parameter dependence of its residual. The implicit-function relation
\begin{equation}
 \frac{\mathrm{d}\bm z_{\rm co}}{\mathrm{d}\bm\theta}=-\left(J-I\right)^{-1}\frac{\partial\bm r}{\partial\bm\theta},\qquad J=\frac{\partial\mathcal M}{\partial\bm z}\Big|_{\rm co},
 \label{eq:ift}
\end{equation}
then gives the sensitivity of the converged fixed point. Tunes, chromaticity, amplitude detuning obtained by numerical analysis of fundamental frequencies (NAFF)~\cite{laskar1993}, and the orbit-response matrix can also be differentiated.

For the injector component, \texttt{track\_differentiable} builds an autograd graph from the field evaluation, particle advance, space-charge solve and beam statistics over a fixed step schedule. Supported parameters include cavity phase and amplitude, solenoid field, quadrupole gradient, charge and laser spot size. Reverse-mode differentiation gives the gradient of a scalar output with respect to all selected parameters. A full Jacobian requires a reverse sweep for each output; the ring Jacobian benchmark uses this procedure. Checkpointing stores selected intermediate states and recomputes the others during the reverse sweeps to reduce memory use. Aperture losses, emission timing and the step at which the cathode image is switched off are not differentiated.

\subsection{Execution and floating-point precision}

A ring element or injector step can execute directly in \textsc{PyTorch}, through \texttt{torch.compile} with Triton kernels~\cite{ansel2024,tillet2019}, or by replaying a CUDA graph. Compilation is applied to reusable element or step functions. Varying scalar settings, including offsets, tilts, screen position and RF phase, are passed as scalar tensors so that scans can reuse the compiled calculation. Injector mesh sizes are drawn from fixed sets: longitudinal cell counts for the cylindrical solver, and a geometric ladder of cell sizes with rungs about 4\% apart for the Cartesian solver. The mesh grows as soon as the bunch outgrows it, but shrinks only when the required size falls below 70\% of the next smaller cell count (cylindrical) or by more than two rungs of about 4\% (Cartesian). This stops a bunch extent that fluctuates about a boundary from triggering a solver rebuild and recompilation at every field update.

For rings at fixed reference energy, a double-single representation stores each coordinate as a pair of float32 values, providing an effective mantissa of roughly 44 bits on a consumer GPU. The benchmarks compare this option with ordinary float32 and float64 and distinguish repeated-call execution from compilation and other initialisation costs.

\subsection{AI-assisted development and verification}
\label{sec:ai-development}

The implementation, tests and documentation, together with the benchmark adapters, analysis scripts and figure-production code, were written predominantly using Anthropic's Claude Sonnet 5.5 through Claude Code and OpenAI's GPT Codex 6.1 Sol through Codex. Some code was written by humans. The authors selected the physical models and reference implementations, set acceptance tolerances, directed the development and reviewed the code and numerical comparisons. The workflow combined literature research, a written feature plan, implementation and quantitative checks against established codes or closed-form results. The authors retain responsibility for the scientific claims and interpretation.

The assistants were used to explore implementation choices as well as to write the adopted code. For instance, a \textsc{JAX}-based injector prototype was implemented and compared with \textsc{PyTorch}, but was not adopted as it showed no benefits. They also helped build adapters to ten external simulation codes across the benchmark cases (Sec.~\ref{sec:setup}), including the ring comparisons in Sec.~\ref{sec:ring}, and to automate photoinjector sensitivity campaigns involving hundreds of runs (Sec.~\ref{sec:clara}). Models were implemented from primary sources and checked against closed-form results or established codes. These examples document the scope of the development and validation work.

AI tools also assisted with manuscript drafting, revision, plots and the workflow illustration, under human supervision. The numerical figures use the saved benchmark data. Automated tests, independent code comparisons and convergence studies assess the resulting implementation on the reported cases. Agreement between AI-generated code and its own generated tests alone does not establish physical correctness. Development took about 12 weeks, with roughly 100--150~h of working sessions (including simulation run times) by the lead author. Claude Code records token use that would cost about \$400 at Anthropic's list API prices for Claude Sonnet 5.5, with Codex usage a small fraction of this.

\section{Benchmark setup}
\label{sec:setup}

\begin{table*}[t]
\caption{Common benchmark cases. The CLARA injector, optimisation and additional validation studies use the separate protocols described with their results.}
\label{tab:cases}
\footnotesize
\setlength{\tabcolsep}{3pt}
\renewcommand{\arraystretch}{1.1}
\begin{tabular}{@{}lllll@{}}
\toprule
\pcell{1.7cm}{Case} & \pcell{4.3cm}{Workload} & \pcell{4.5cm}{Observables} & \pcell{3.0cm}{Reference} & \pcell{3.5cm}{Tolerance} \\
\midrule
\pcell{1.7cm}{\texttt{ring\_tunes}} & \pcell{4.3cm}{48~m double-bend-achromat ring, 629~MeV electrons; 5 particles at $x=1,2,4,6,8$~mm ($y=1$~mm); 1024 turns} & \pcell{4.5cm}{Fractional tunes (Hann-windowed frequency analysis); first-turn residuals of $(x,p_x,y,p_y)$} & \pcell{3.0cm}{\textsc{elegant}, order 6 with 16/32 slices} & \pcell{3.5cm}{Tune $10^{-5}$; first turn $10^{-8}$ in each component} \\
\midrule[0.2pt]
\pcell{1.7cm}{\texttt{ring\_grad}} & \pcell{4.3cm}{Same ring; one particle ($x=4$~mm, $y=1$~mm); state after 39 turns; knobs $k_1$ of three quadrupole families and $k_2$ of two sextupoles} & \pcell{4.5cm}{$4\times5$ Jacobian; finite differences (FD) at relative steps $10^{-2}$--$10^{-5}$ with Richardson extrapolation; automatic differentiation (AD) where available} & \pcell{3.0cm}{\textsc{elegant} central-difference Jacobian} & \pcell{3.5cm}{Unit-invariant limits of Sec.~\ref{sec:grad}} \\
\midrule[0.2pt]
\pcell{1.7cm}{\texttt{sc\_drift}} & \pcell{4.3cm}{20~pC, 5~MeV kinetic energy electron bunch, 3~m free-space drift, 20\,000 macroparticles, 20 stations} & \pcell{4.5cm}{rms size, rms length, normalised emittance, rms energy spread, mean-energy gain} & \pcell{3.0cm}{\textsc{ASTRA} 3-D solver, $32^3$ mesh (finer meshes in Sec.~\ref{sec:drift})} & \pcell{3.5cm}{Size and length 3\%, emittance 5\%} \\
\midrule[0.2pt]
\pcell{1.7cm}{\texttt{linear\_fodo}} & \pcell{4.3cm}{1~m hard-edge FODO, 3~MeV protons, $k_1=\pm32$~m$^{-2}$} & \pcell{4.5cm}{Position and normalised momentum} & \pcell{3.0cm}{Closed-form thick-quadrupole matrices} & \pcell{3.5cm}{$10^{-11}$ in each unit} \\
\bottomrule
\end{tabular}
\end{table*}

Table~\ref{tab:cases} defines four common benchmark cases. Eleven codes participate where their models and interfaces support the calculation: \textsc{Slipstream}, \textsc{elegant}, \textsc{Xsuite}, \textsc{Cheetah}, \textsc{JuTrack}, \textsc{SciBmad}, \textsc{ASTRA}, \textsc{GPT}, \textsc{ImpactX}, \textsc{Ocelot}~\cite{agapov2014} and \textsc{HELIX}. Table~\ref{tab:versions} records their versions. The CLARA injector and later validation studies have separate protocols, stated with those results.

Nominal strengths and slice counts alone do not define an identical map across codes. \textsc{SciBmad}'s default tracking method includes hard-edge fringe fields, so the comparison selects its symplectic integrator for quadrupoles and sextupoles, with the fringes off and exact sector bends. \textsc{Cheetah}'s nonlinear setting combines drift--kick--drift transport with a second-order sextupole. The \textsc{Xsuite} comparison explicitly selects its exact drift and sextupole models. The ring study therefore checks individual elements before interpreting tune and derivative differences.

Adapters convert coordinates explicitly, including \textsc{Cheetah}'s energy coordinate, \textsc{elegant}'s transverse slopes and the common-time bunches of \textsc{ASTRA} and \textsc{GPT}. Position and normalised-momentum residuals are reported separately. Where a combined error is used, each component is first divided by its stated physical scale.

\begin{table}[t]
\caption{Versions of the codes. All run natively on Windows~11 unless a platform is given; \textsc{PyTorch} 2.14.0 with CUDA 13.2 (2.14.1 for \textsc{HELIX}), NVIDIA driver 616.56.}
\label{tab:versions}
\footnotesize
\setlength{\tabcolsep}{3pt}
\begin{tabular}{@{}llp{3.2cm}@{}}
\toprule
Code & Version & Notes \\
\midrule
\textsc{Slipstream} & 1.0.1 & \textsc{Warp} 1.17.0 \\
\textsc{elegant} & 2026.3.0 & serial; \textsc{Pelegant} with MS-MPI \\
\textsc{Xsuite} & \textsc{xtrack} 0.111.6 & \textsc{xfields} 0.27.2, \textsc{xobjects} 0.6.10; WSL2/Ubuntu \\
\textsc{Cheetah} & 0.8.4 & \\
\textsc{JuTrack} & 1.5.0, commit 5b5eebd & \textsc{Julia} 1.10.4 \\
\textsc{SciBmad} & 0.6.0 & BeamTracking 0.8.7, CUDA.jl 6.4.1; \textsc{Julia} 1.12.7 \\
\textsc{ASTRA} & 4.0 (DESY, 2022) & \\
\textsc{GPT} & build of 14 October 2024 & \\
\textsc{ImpactX} & 26.10 & WSL2/Ubuntu, CPU \\
\textsc{Ocelot} & 26.6.1 & WSL2/Ubuntu \\
\textsc{HELIX} & 1.20.0 & \\
\bottomrule
\end{tabular}
\end{table}

All simulations and benchmarks use one workstation: an AMD Ryzen 9 7950X with 16 cores and an NVIDIA RTX 3050 desktop GPU with 8~GiB, running Windows 11 with WSL2 where specified. The common-code ring, Jacobian and drift benchmarks use one CPU thread per CPU code. Each common benchmark setting was run in three independent processes in shuffled order. The reported warm time is the median, over processes, of the repeated-call time.

Repeated-call times exclude imports, model construction and compilation. They include the adapter work required to produce the common output and, for external executables, process launch and file input and output. They measure the latency of this workflow rather than isolated kernel execution. The tolerances in Table~\ref{tab:cases} provide acceptance levels for the selected tests; errors are reported throughout.

The later studies use explicitly identified exceptions: CLARA timings include multithreaded \textsc{ASTRA} and \textsc{GPT}, the CSR comparison includes a multithreaded \textsc{Slipstream} run, and the slowest drift scaling points are single runs. The injector optimisation study uses concurrent multithreaded processes and is reported under that resource allocation. These protocols are specified beside the relevant tables and figures.

For the work carried out and benchmarked in this paper we use \textsc{ASTRA} and \textsc{elegant} as the baseline against which \textsc{Slipstream} is compared. This is a practical choice based on the standard codes and workflows in use within the Accelerator Science and Technology Centre (ASTeC), and the existing software framework used for accelerator design and simulations of several facilities.

\section{Benchmark results}
\label{sec:results}

\subsection{Nonlinear ring tunes and element models}
\label{sec:ring}

The ring comparison tests amplitude-dependent tunes and the first-turn map. Amplitude-dependent tunes depend on the nonlinear terms of the element maps, so they are sensitive to differences between element models. We then compare each code's single-element maps with an independent integration, to determine whether differences arise from the element models or from the tracking framework.

\begin{table*}[t]
\caption{Ring tunes and first-turn map against \textsc{elegant} (order 6, 16/32 slices). First-turn entries are maximum absolute residuals, separately over $(x,y)$ in metres and $(p_x,p_y)$ in normalised units. Configuration tags give integration order (o), quadrupole/sextupole slices (k or s), and bend slices (b); for \textsc{Slipstream}, q and s give the steps per quadrupole and per sextupole; yos denotes Yoshida-4 in \textsc{Xsuite}, n denotes \textsc{JuTrack} steps, and for \textsc{SciBmad} o and k give the symplectic order and the steps per quadrupole and sextupole (hard-edge fringes off, exact bends; ``default'' is the shipped method, which includes fringe fields). Entries within the tolerance are in bold. Warm times are medians of three processes; the spread between processes is 1.2\% or less for \textsc{Slipstream} on the GPU and at most 12\% for the other settings.}
\label{tab:ring}
\footnotesize
\setlength{\tabcolsep}{4pt}
\begin{tabular}{@{}llrccc@{}}
\toprule
Code & Setting & Warm [s] & Tune error & First turn $\max|\Delta x,\Delta y|$ [m] & First turn $\max|\Delta p_x,\Delta p_y|$ \\
\midrule
\textsc{Slipstream} & GPU f32, o6q8s1 & 0.482 & $\bm{1.75\times10^{-6}}$ & $5.04\times10^{-8}$ & $3.03\times10^{-8}$ \\
\textsc{Slipstream} & GPU f64, o6q8s1 & 2.07 & $\bm{1.66\times10^{-9}}$ & $\bm{1.10\times10^{-11}}$ & $\bm{2.68\times10^{-11}}$ \\
\textsc{Slipstream} & GPU f64, o6k16b32 & 5.63 & $\bm{3.59\times10^{-11}}$ & $\bm{5.04\times10^{-12}}$ & $\bm{3.57\times10^{-12}}$ \\
\textsc{Slipstream} & GPU f32 double-single, o6q8s1 & 2.13 & $\bm{2.8\times10^{-9}}$ & $\bm{1.3\times10^{-10}}$ & $\bm{5.3\times10^{-11}}$ \\
\textsc{Slipstream} & CPU f64 compiled, o6q8s1 & 4.57 & $\bm{1.69\times10^{-9}}$ & \multicolumn{2}{c}{---} \\
\textsc{elegant} & CPU f64, o6s8b16 & 0.701 & $\bm{3.82\times10^{-9}}$ & $\bm{3.32\times10^{-10}}$ & $\bm{2.55\times10^{-10}}$ \\
\textsc{Xsuite} & CPU f64, yosk8b16, exact model & 0.199 & $\bm{6.86\times10^{-6}}$ & $5.81\times10^{-8}$ & $1.22\times10^{-7}$ \\
\textsc{JuTrack} & CPU f64, n32b64 & 0.853 & $4.17\times10^{-5}$ & $2.39\times10^{-5}$ & $7.10\times10^{-7}$ \\
\textsc{SciBmad} & CPU f64, o4k4 & 0.898 & $\bm{8.86\times10^{-9}}$ & $\bm{7.48\times10^{-11}}$ & $\bm{1.31\times10^{-10}}$ \\
\textsc{SciBmad} & CPU f64, o6k32 & 4.77 & $\bm{8.08\times10^{-11}}$ & $\bm{5.26\times10^{-12}}$ & $\bm{4.04\times10^{-12}}$ \\
\textsc{SciBmad} & CPU f64, default (fringes) & 0.745 & $3.96\times10^{-3}$ & $4.37\times10^{-5}$ & $5.50\times10^{-5}$ \\
\textsc{Cheetah} & CPU f64, drift-kick-drift, 16 steps & 138 & $1.04\times10^{-4}$ & $1.50\times10^{-7}$ & $1.86\times10^{-6}$ \\
\textsc{Cheetah} & CPU f64, linear map & 1.35 & $3.36\times10^{-3}$ & $5.26\times10^{-4}$ & $2.03\times10^{-4}$ \\
\bottomrule
\end{tabular}
\end{table*}

\begin{figure*}[t]
\includegraphics[width=\textwidth]{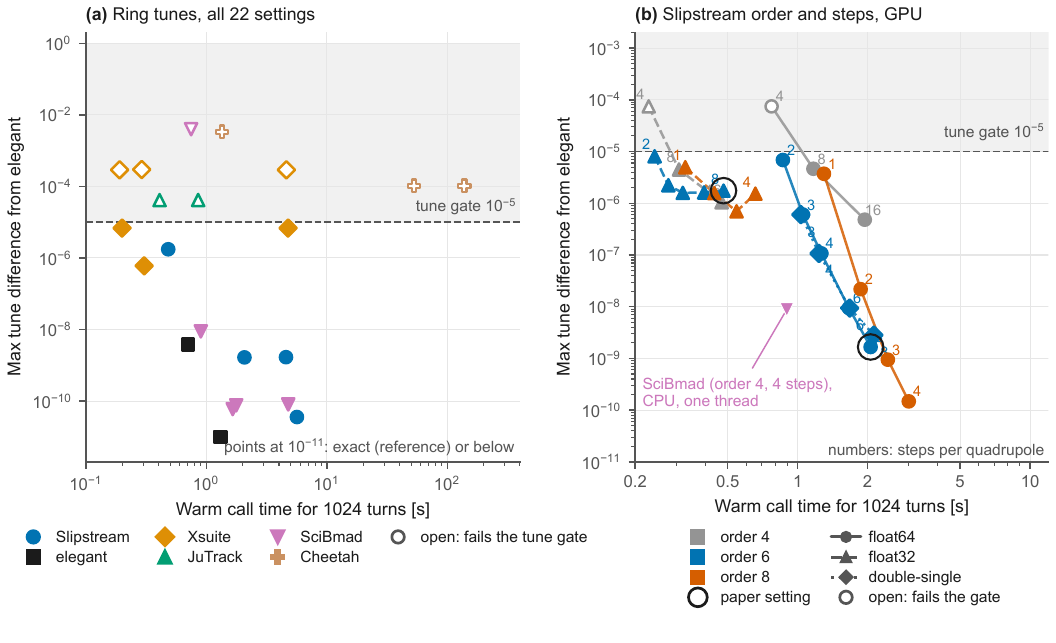}
\caption{Nonlinear ring tunes. (a) Largest tune difference from \textsc{elegant} over five launches and both planes, against warm time for 1024 turns, for all 22 settings; open markers exceed the $10^{-5}$ tolerance. (b) Largest tune difference for \textsc{Slipstream} on the GPU against warm time for 1024 turns, for varying integration orders in float32, float64 and double-single (hybrid) arithmetic. Labels give the quadrupole steps, with one step per sextupole and exact sector bends. The rings mark the setting adopted here (order 6, eight steps per quadrupole).}
\label{fig:ring}
\end{figure*}

\textsc{Slipstream}'s float64 tunes agree with \textsc{elegant} to $10^{-9}$ for five particle launches of increasing amplitude (Table~\ref{tab:ring}, Fig.~\ref{fig:ring}a). Refining the slicing reduces the largest tune difference to $4\times10^{-11}$; \textsc{elegant} at 8/16 slices differs from the 16/32-slice reference run by $4\times10^{-9}$. \textsc{Slipstream}'s first-turn residuals are a few $10^{-11}$, reported separately for positions and normalised momenta. \textsc{Xsuite} with explicitly selected exact drift and sextupole models also meets the tune tolerance.

The nonlinear settings of \textsc{Cheetah} and \textsc{JuTrack} differ from the reference by $4\times10^{-5}$ to $1\times10^{-4}$ despite increased slicing (\textsc{Cheetah}'s faster linear model has no amplitude detuning and differs by $3\times10^{-3}$). \textsc{SciBmad}, with its hard-edge fringes off and exact sector bends, meets the tolerances at order 4 with four steps per magnet ($9\times10^{-9}$) and converges to $8\times10^{-11}$; its default settings apply hard-edge fringe fields and differ by $4\times10^{-3}$, which is a modelling difference rather than an integration error.

Precision changes which observables meet the stated tolerances. GPU float32 meets the tune tolerance but not the first-turn tolerance. The double-single hybrid method, storing each coordinate as two float32 values, meets both at about the float64 cost for this small workload. In the separate linear-limit check, \textsc{Slipstream}'s order-6 hard-edge proton FODO agrees with the analytic transport to round-off in positions in metres and in normalised momenta.

We tested the element-model explanation for the tune differences by comparing each code's single-element maps with an independent integration of the corresponding Hamiltonian on 135 sampled states, with $x$ up to 8~mm and $\delta$ up to 1\%. \textsc{Slipstream}'s largest absolute component residual is $2\times10^{-13}$ and \textsc{elegant}'s is about $10^{-9}$, with position residuals measured in metres and momenta in normalised units. \textsc{SciBmad}'s four maps agree with the reference to $5\times10^{-16}$. The other codes have one or two maps with residuals between $10^{-7}$ and $10^{-4}$ (\textsc{JuTrack}'s bend, the quadrupole and sextupole of \textsc{Cheetah} and \textsc{Xsuite}'s default adaptive drift) which adequately explains why they miss the tune gate in this scenario.

Figure~\ref{fig:ring}(b) shows how the accuracy of \textsc{Slipstream}'s ring tracking depends on its integration order and step count. The three orders follow one curve of tune difference against warm time, so a higher order does not lower the cost of a given accuracy. Assigning the steps to the quadrupoles, with one step per sextupole, is cheaper: eight steps per quadrupole (ringed) gives the same tune difference as eight steps in every magnet in about 30\% less time in both precisions. In float32 the tune difference stops improving near $1.6\times10^{-6}$ from four steps per quadrupole, so further steps add only cost. In float64, six steps per quadrupole reach the tune difference of \textsc{SciBmad}'s four-step setting in about 65\% more time than \textsc{SciBmad} takes on one CPU thread. In this small example the comparison measures calculation efficiency, not scaling efficiency, which is shown in the next section.

\subsection{Cost and scaling of ring tracking}
\label{sec:ringcost}

A GPU gives little benefit for the five-particle tune test, where dispatch latency dominates. \textsc{Xsuite} takes 0.2~s on one CPU thread for 1024 turns, compared with 0.5~s for \textsc{Slipstream} float32 and 2.1~s for float64 on the GPU. \textsc{SciBmad} (order 4, four steps per quadrupole and sextupole) takes 0.9~s on one CPU thread and 4.8~s in float32 or 15~s in float64 on the GPU, where launch latency also dominates.

\begin{figure*}[t]
\includegraphics[width=\textwidth]{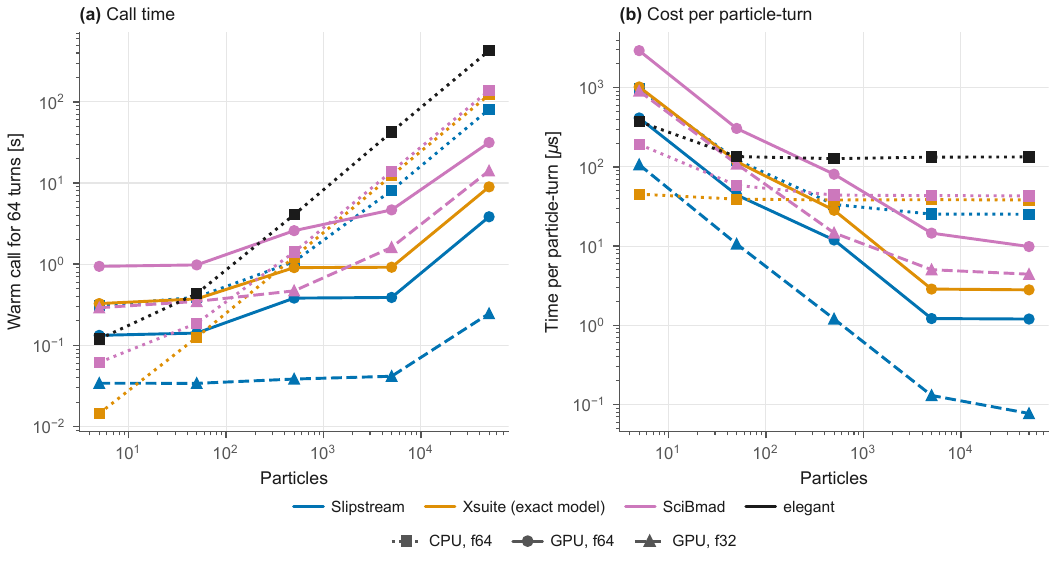}
\caption{Particle-count scaling of the ring workload (64 turns). (a) Warm call time; (b) time per particle-turn. }
\label{fig:scaling}
\end{figure*}

Figure~\ref{fig:scaling} tests larger ensembles, from 5 to 50\,000 particles over 64 turns. GPU runtime changes little up to several thousand particles, while CPU runtime increases approximately linearly with particle count. \textsc{Slipstream}'s GPU calculation becomes faster than single-thread \textsc{Xsuite} from 50 particles in float32 and 500 in float64. At 5000 particles it is 297 and 32 times faster, respectively. \textsc{Xsuite} on the same GPU in float64 takes 2.3--2.6 times as long as \textsc{Slipstream} at every particle count, and is 13--14 times faster than its own single-thread CPU calculation at 5000 and $50\,000$ particles. Equal precision does not make the element maps identical, so the map and tune checks still govern the comparison. \textsc{SciBmad}, with order 4 and four steps per quadrupole and sextupole, takes 0.06~s for five particles on one CPU thread, rising linearly to 138~s at $50\,000$ particles. On the GPU it takes 0.9~s (float64) and 0.3~s (float32) for five particles and overtakes its own CPU calculation at about 1000 particles in float64 and about 100 in float32. At $50\,000$ particles it is 8.2 and 57 times slower than \textsc{Slipstream} in float64 and float32, and 3.5 times slower than \textsc{Xsuite} in float64. Its float32 results also do not meet the tune tolerance ($1.5\times10^{-4}$ over 1024 turns).

At 50\,000 particles, \textsc{Slipstream} takes 0.077~$\mu$s per particle-turn in float32 and 1.2~$\mu$s in float64 on the GPU, and \textsc{Xsuite} 2.8~$\mu$s in float64. Compiled \textsc{Slipstream} and \textsc{Xsuite} on one CPU thread take 25 and 38~$\mu$s, and \textsc{elegant} takes 134~$\mu$s; \textsc{SciBmad} takes 43~$\mu$s on one CPU thread and 9.9 and 4.4~$\mu$s on the GPU in float64 and float32. These ratios compare the stated GPU with one CPU thread and exclude initialisation; \textsc{Xsuite}'s GPU line build takes about 5~min per process and \textsc{SciBmad}'s first GPU call about 40~s.

\subsection{Ring derivatives and optics matching}
\label{sec:grad}

\begin{figure*}[t]
\includegraphics[width=\textwidth]{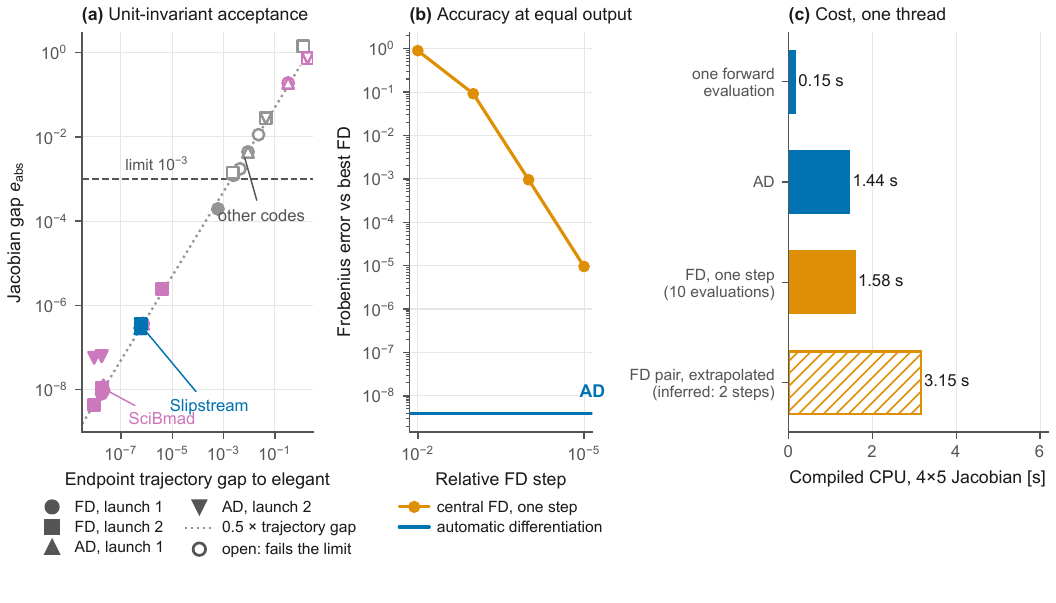}
\caption{Derivatives of the tracked state with respect to five magnet strengths. (a) Dimensionless Jacobian difference $e_{\rm abs}$ against each code's scaled endpoint-trajectory difference, both measured from \textsc{elegant}, for two launches and knob sets (circles and squares: finite differences; triangles: AD). Filled markers are within the limit. (b) Frobenius error of one central finite-difference step against the best extrapolated finite-difference Jacobian, and of AD (horizontal line). (c) Time for one Jacobian on compiled CPU, one thread; the hatched bar is estimated as twice the single-step cost.}
\label{fig:grad}
\end{figure*}

Before comparing derivatives between codes, we check each derivative against the same code's finite differences (FD). \textsc{Slipstream}'s AD and central-FD $4\times5$ Jacobians agree to a relative Frobenius error of $4\times10^{-9}$. Compiling the reusable CPU calculations reduces the AD time to 1.5~s and the FD time to 1.8~s, on one thread. These are 52 and 25 times faster than direct, uncompiled execution; the results change by less than $2\times10^{-13}$.

A cross-code Jacobian comparison must also allow for differences in the underlying trajectories. We scale each derivative to a dimensionless sensitivity $S_{ij}=\theta_j(\partial y_i/\partial\theta_j)/\sigma_i$, where $\theta_j$ is the nominal magnet strength and $\sigma_i$ is the rms amplitude of output coordinate $i$ over the first 40 turns, with $i$ one of $(x,p_x,y,p_y)$. The on-momentum case uses \textsc{elegant}'s trajectory for these scales; the second, off-momentum case uses the fine native trajectory. This scaling prevents a change of position or momentum units from changing the judgement.

We apply two limits relative to \textsc{elegant}'s FD Jacobian. The largest absolute sensitivity difference, divided by the largest reference sensitivity, must satisfy $e_{\rm abs}\le10^{-3}$. Among reference entries larger than $10^{-3}$ of that largest sensitivity, the worst relative difference must satisfy $e_{\rm rel}\le1\%$. \textsc{Slipstream} meets both limits by more than three orders of magnitude on two launches, including one with $\delta=3\times10^{-3}$ and a different parameter set; $e_{\rm abs}\approx3$--$4\times10^{-7}$. \textsc{SciBmad}, with its fringes off and exact bends, also meets both limits: its forward-mode AD (\textsc{ForwardDiff}) agrees with its own FD to $4\times10^{-9}$, and differs from \textsc{elegant} by $e_{\rm abs}=3.7\times10^{-7}$ at order 4 with four steps and $1.3\times10^{-8}$ at order 6 with 32 steps on the first launch ($2.4\times10^{-6}$ and $6.3\times10^{-8}$ on the second). For one particle over 39 turns it computes the $4\times5$ Jacobian in 0.09--0.30~s, 2.7 to 4.4 times faster than one central-FD step.

The scaled Jacobian differences of the other codes, and of \textsc{SciBmad} with its shipped fringe-field model, range from $10^{-3}$ to 1 and are comparable to their scaled endpoint-trajectory differences [Fig.~\ref{fig:grad}(a)]. After 39 turns, a tune offset of $10^{-4}$ already gives a phase difference of about 0.025~rad. The cross-code derivative differences therefore largely follow the element-model differences. Agreement with a code's own FD calculation remains the direct check of its derivative implementation.

Finite-difference accuracy also depends on step size. The relative Frobenius error against the best extrapolated FD result is 0.09 at a relative step of $10^{-3}$, $10^{-3}$ at $10^{-4}$ and $10^{-5}$ at $10^{-5}$. AD gives $4\times10^{-9}$ [Fig.~\ref{fig:grad}(b)]. Extrapolating a pair of steps improves FD accuracy at approximately twice the cost of a single step.

\begin{figure*}[t]
\includegraphics[width=\textwidth]{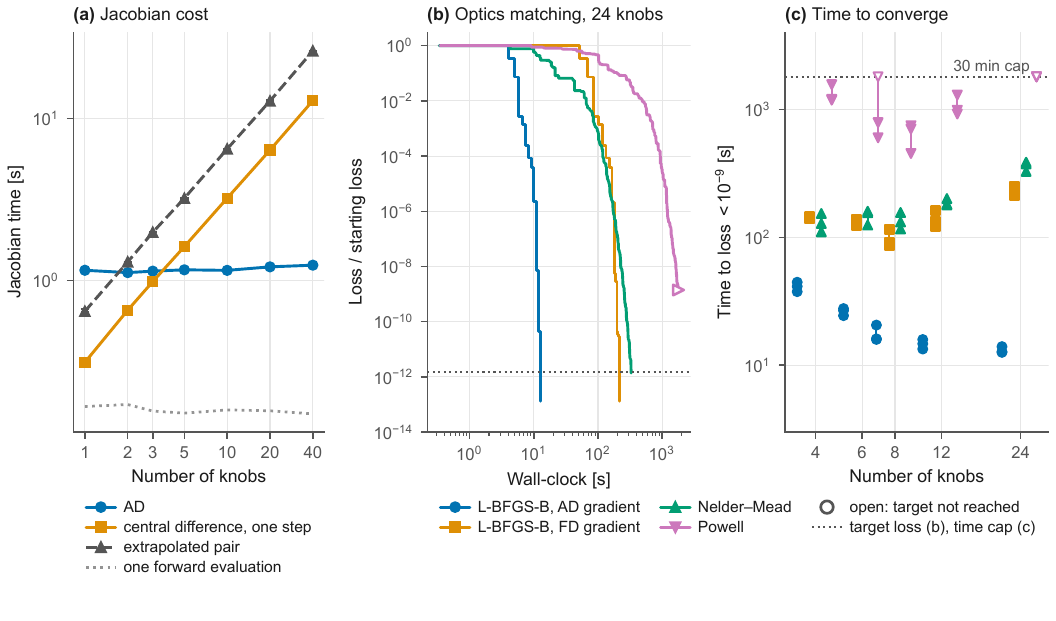}
\caption{Cost of derivatives and of optimisation with them. (a) Time of Jacobian calculation of the four outputs $(x,p_x,y,p_y)$ against the number of knobs, for AD, for central differences at one relative step of $10^{-5}$ and for an extrapolated pair of steps (compiled CPU, one thread, medians of three processes with the observed range; dotted: one forward evaluation). (b) Optics matching of a perturbed 48~m ring with all 24 quadrupoles free: loss relative to its starting value against elapsed time, which includes every objective evaluation; the open marker is the 30-minute cap. (c) Time to a loss below $10^{-9}$ for every case and optimiser, against the number of knobs (open markers: cap reached).}
\label{fig:adcost}
\end{figure*}

\paragraph*{Cost as parameters are added.} For the four-output ring Jacobian, AD takes 1.1--1.2~s across the tested range of 1--40 magnet-strength parameters, about seven forward evaluations [Fig.~\ref{fig:adcost}(a)]. The implementation uses one reverse sweep per output; adding input parameters leaves the number of sweeps fixed. Central differences require two forward evaluations per parameter.

A single central-FD step of $10^{-5}$ takes 0.3~s for one parameter and 13~s for 40. The extrapolated pair takes 0.6 and 26~s. AD becomes faster at three to five parameters for a single step and at two for the pair; at 40 parameters it is 10 and 21 times faster. Both FD procedures satisfy $e_{\rm abs}\le1.5\times10^{-6}$, below the acceptance limit of $10^{-3}$. The timing comparison is therefore between methods meeting the required tolerance, although their achieved errors differ.

AD also supports parameters whose nominal value is zero. For quadrupole offsets it agrees with central FD to $5\times10^{-9}$ at a 0.1~$\mu$m step. The AD/FD crossover measured here only applies to this four-output ring calculation. The ten-to-twelve-parameter crossover reported for a PIC cycle in Ref.~\cite{pathak2026} concerns a different system and workload, and is difficult to directly compare.

\paragraph*{Optics matching.} We next test whether the derivative saving reduces optimisation time. Random 1\% quadrupole-gradient errors perturbed the 48~m ring. The objective restores both tunes and the beta functions at the observation point:
\begin{equation}
 L_{\rm match}=\sum_{j=x,y}\left[\left(\frac{\Delta Q_j}{10^{-3}}\right)^2+\left(\frac{\beta_j/\beta_{j,0}-1}{10^{-2}}\right)^2\right],
 \label{eq:matching-loss}
\end{equation}
where subscript 0 denotes the unperturbed target and $\Delta Q_j$ is the tune difference. These optimisation runs use order-4 maps with 4 quadrupole/sextupole slices and 8 bend slices. Each run uses one CPU thread, with all objective evaluations and reverse sweeps included in elapsed time. We compare L-BFGS-B using AD or central-FD gradients with derivative-free Nelder--Mead and Powell methods [Fig.~\ref{fig:adcost}(b,c)].

With all 24 quadrupoles adjustable, an exact correction exists. AD reaches $L_{\rm match}\le10^{-9}$ in 13--14~s across three seeds. L-BFGS-B with FD takes 214--251~s and Nelder--Mead 328--387~s. Powell reaches the 30-minute cap without meeting the target in all three runs performed.

For 4--12 parameters, the imposed errors are shared by the corresponding blocks of neighbouring quadrupoles so that an exact correction again exists. AD converges in 13--45~s in all 12 cases; FD and Nelder--Mead take 3--12 times longer. The advantage over FD grows from 3.4 at four parameters to 18 at 24.

\paragraph*{Ring tolerance budget.} The same four outputs, the two tunes and the two beta functions at the observation point, give a first-order tolerance budget for the 24 quadrupoles. We assign independent Gaussian gradient errors of 0.1\% to each quadrupole and obtain the budget from the AD Jacobian, from the central-FD Jacobian (relative step $10^{-5}$) and from a Monte Carlo of 40 machines, on one CPU thread with the order-4 maps of the matching runs. One forward evaluation takes 0.78~s. The AD Jacobian takes 2.2~s, about 2.8 forward evaluations, and stays between 2.0 and 2.2~s from 1 to 24 parameters. Central FD takes 41~s for 24 parameters and the Monte Carlo 35~s, so AD is 19 and 16 times faster. The AD and FD Jacobians agree to a relative Frobenius error of $4\times10^{-9}$.

The linear prediction of the output standard deviations lies within 6\% of those from a 400-machine Monte Carlo at 0.1\%, 0.3\% and 1\% errors; the sampling uncertainty of that reference is about 3.5\%. The 40-machine estimate alone scatters between 0.74 and 1.23 of the reference. The Jacobian also attributes the variance: two quadrupoles of one family contribute 47\% of the $\beta_x$ variance. These outputs are linear-optics quantities that respond almost linearly to the errors, so this is a favourable case, and the Monte Carlo here is a serial calculation; an ensemble batched on the GPU would cost less.

\subsection{Injector derivatives and memory}
\label{sec:injgrad}

An injector derivative study used the 5~pC CLARA input deck, separately from the 20--250~pC accuracy study in Sec.~\ref{sec:clara}. It checks differentiation through the field-map advance and cylindrical space-charge solve over the full 3.37~m line on CPU in float64, with 1024 macroparticles and a 0.5~ps step. The cathode image is included, as in the 20--250~pC accuracy study. The differentiable calculation takes the image as a number of steps, 379 (190~ps) here, set from the nominal run, which drops the image once the bunch centroid is 5~cm from the cathode. Like the step schedule and the emission times, this number is held fixed, so the derivatives do not include the shift of the switch-off step with the parameters. Six parameters vary: the phase and field of each cavity, the gun solenoid field and bunch charge. The four outputs are horizontal emittance, rms size, rms bunch length and longitudinal emittance.

\begin{figure*}[t]
\includegraphics[width=\textwidth]{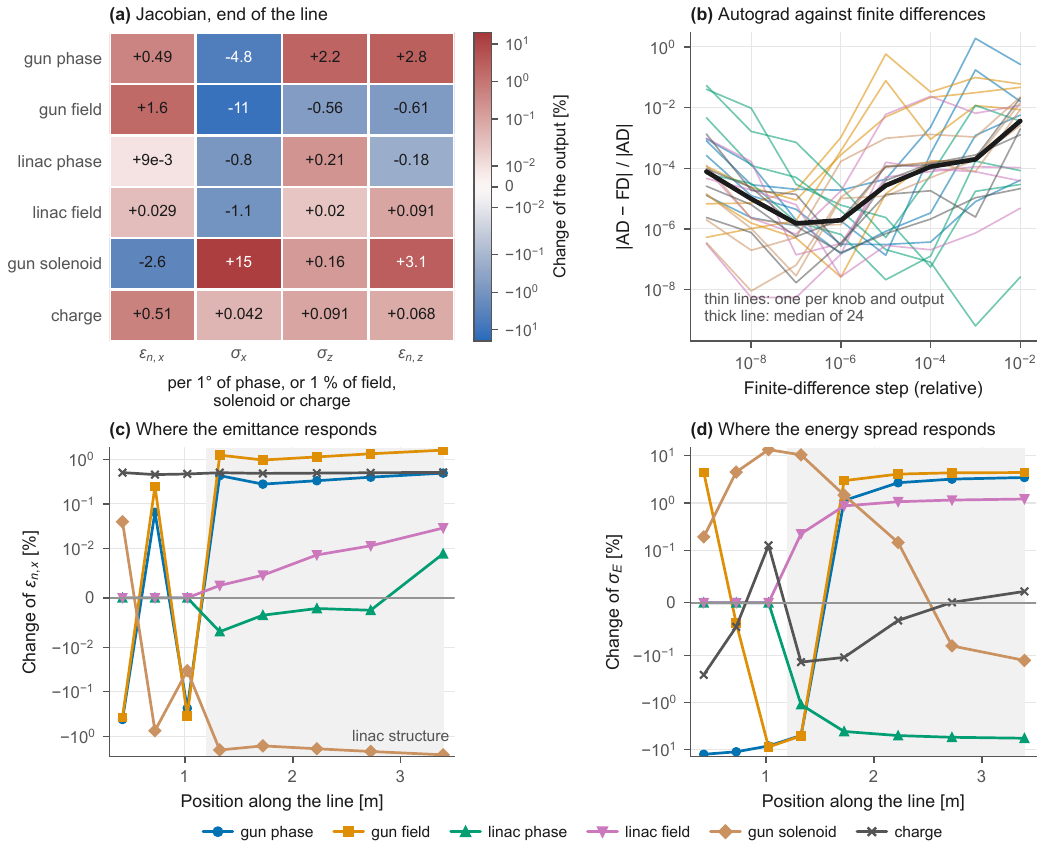}
\caption{Derivatives for the 5~pC CLARA injector. (a) Four-output Jacobian for six parameters, expressed as percentage output changes for $1^\circ$ phase or 1\% field or charge changes. (b) Relative AD--FD differences over the tested step sizes, with one curve per parameter--output pair and the median. The derivative checks use CPU float64, 1024 particles, a 0.5~ps step and the cathode image. (c, d) Emittance and rms energy-spread sensitivities at planes through the beamline; shading marks the first linac.}
\label{fig:injad}
\end{figure*}

The 24 derivatives (four outputs by six parameters) agree with central FD to relative errors between $6\times10^{-10}$ and $2\times10^{-5}$ at their best steps (Fig.~\ref{fig:injad}(a,b)); AD obtains them from four reverse sweeps, one per output. The best FD step differs between sensitivities: the calculated outputs have relative fine structure of up to about $10^{-5}$ as parameters vary. A larger FD step averages over that structure, whereas AD returns the local slope of the implemented calculation.

\paragraph*{Where the beam responds.} Evaluating sensitivities at intermediate planes shows where the final response develops (Fig.~\ref{fig:injad}c,d). Before the accelerating structure, energy spread responds to the gun phase and field. After acceleration, it also responds to the structure phase, reaching about $-5.7\%$ per degree at the line exit. The exit energy-spread derivatives agree with central FD to better than $10^{-6}$ for all six parameters.

Emittance is most sensitive to solenoid field and gun field. After acceleration, a 1\% increase in solenoid field changes it by $-1.6$ to $-2.6\%$, while a 1\% increase in gun field changes it by $+1.0$ to $+1.6\%$ along the sampled downstream planes. The structure-phase and structure-field sensitivities are below 0.01\% per degree and 0.03\% per 1\% field change, respectively. This localisation is consistent with the emittance being largely determined early in the line.

\paragraph*{Memory.} Reverse-mode differentiation with space charge records information from both the particle advances and the field solves, so memory grows with the number of steps, macroparticles and grid cells~\cite{dhamrait2026}. We measured the peak allocation of the float32 forward and reverse calculation on the GPU, with the cathode image. Without checkpointing, the injector calculation needs about 1.8~MiB per step for 4096 macroparticles, measured up to 2200 steps (3.9~GiB), and would exceed the GPU's 8~GiB capacity at about 4700 steps. We did not run longer calculations without checkpointing: beyond the capacity the driver spills the allocation into host memory, so such runs would not represent resident-GPU calculations.

Storing checkpoints about every $\sqrt{n}$ steps and recomputing the intervening states reduces the memory to 0.21~GiB at 10\,000 steps. At 3551 steps the remaining storage grows approximately linearly with particle number, reaching 1.8~GiB at 65\,536 macroparticles; the corresponding run without checkpointing was not made. The derivatives above are taken with the step schedule and the particle emission times held fixed. Emission-timing parameters, including the laser pulse duration, are excluded: each macroparticle is released at the first step after its emission time, so the output changes in jumps rather than smoothly, and its derivative is zero wherever it exists.

\subsection{Injector jitter and tuning}
\label{sec:injapps}

\paragraph*{Tolerance budget.} A Jacobian can propagate parameter jitter to predict uncertainty in beam observables. The jitter calculation uses the same 5~pC, 1024-particle, 0.5~ps setup, with the cathode image, as the derivative check. This calculation uses five outputs: horizontal emittance, rms size, rms bunch length, mean energy and rms energy spread. It replaces longitudinal emittance in the preceding four-output derivative check with mean energy and energy spread. The assumed independent Gaussian errors are $0.1^\circ$ in each RF phase, 0.1\% in each cavity field, 0.01\% in solenoid field and 1\% in charge.

\begin{figure*}[t]
\includegraphics[width=\textwidth]{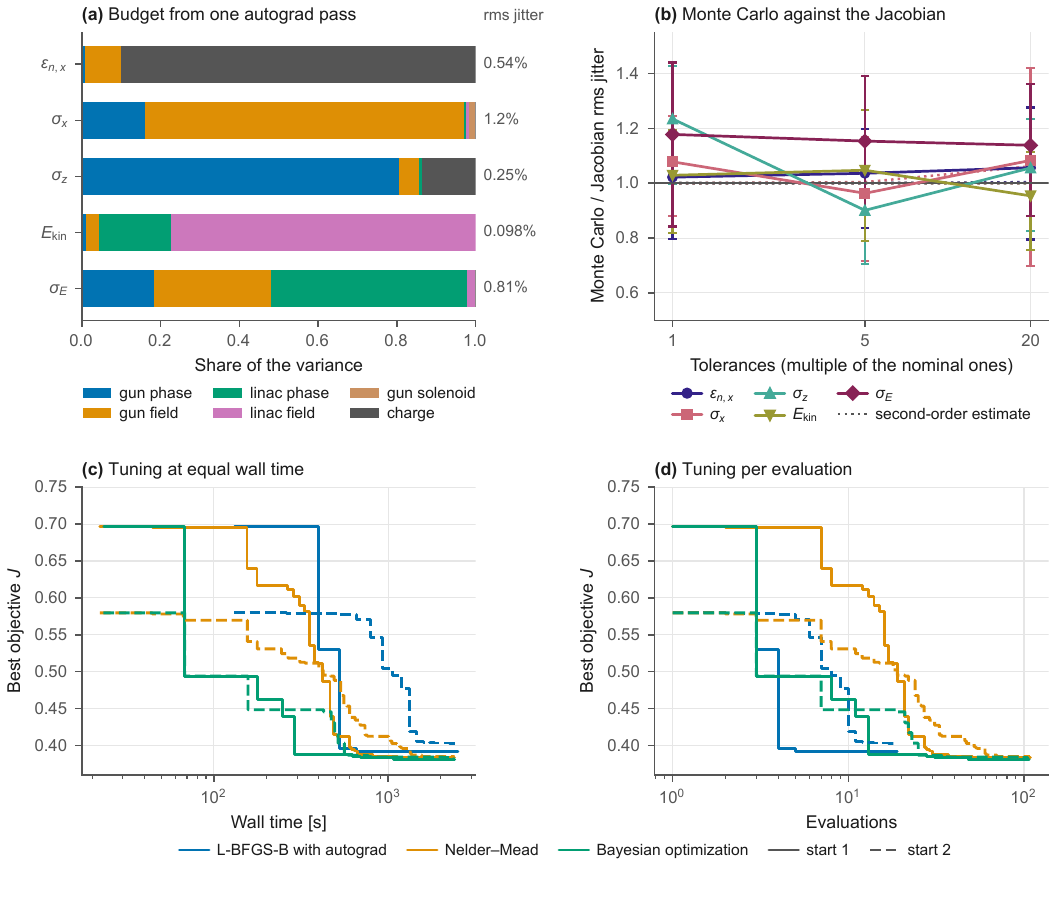}
\caption{Applications of the injector Jacobian. (a) Share of the variance of five end-of-line outputs due to each knob, for independent Gaussian errors of $0.1^\circ$ in each phase, 0.1\% in each cavity field, 0.01\% in the solenoid field and 1\% in the charge, from the five-output Jacobian; the rms jitter is at the right. (b) Ratio of the rms jitter of a Monte Carlo of 40 runs to that predicted from the Jacobian at 1, 5 and 20 times these tolerances, with its 95\% bootstrap interval; dotted lines add the curvature measured at $\pm10$ standard deviations. (c, d) Best objective of three methods tuning five knobs, against wall time and against the number of evaluations, from two starting points (solid and dashed); each method ran for 40~min in a separate five-thread process, concurrently on the same 16-core CPU.}
\label{fig:injapps}
\end{figure*}

The linearised calculation predicts rms output jitters of 0.1--1.2\% (Fig.~\ref{fig:injapps}a). Charge contributes 90\% of the emittance variance, gun field 81\% of the size variance, gun phase 81\% of the bunch-length variance, linac field 77\% of the energy variance and linac phase 50\% of the energy-spread variance.

A Monte Carlo study uses 40 runs at each of 1, 5 and 20 times these tolerances (Fig.~\ref{fig:injapps}b). Its output jitters are consistent with the linearised predictions within the sampling uncertainty at the tested scales. One-at-a-time FD over $\pm10$ standard deviations reproduces the variance budget to 1\%. The five-output autograd Jacobian costs about 23 forward runs, compared with 12 for the one-at-a-time FD calculation and 120 for the Monte Carlo. For these six inputs, FD is the cheaper way to obtain this.

\paragraph*{Tuning.} The Jacobian can supply gradients of objectives defined from the beam observables. We vary five parameters---gun phase and field, solenoid field, laser spot size and structure phase---to minimise a dimensionless objective
\begin{equation}
 L_{\rm inj}=\frac{\epsilon_{n,x}}{1~\mathrm{mm\,mrad}}+0.05\left(\frac{E_{\rm kin}-30~\mathrm{MeV}}{1~\mathrm{MeV}}\right)^2.
 \label{eq:injector-loss}
\end{equation}
The energy penalty makes structure phase relevant as well as the parameters that affect emittance. L-BFGS-B with AD, Nelder--Mead and Bayesian optimisation each receive 40~min from each of two detuned starts (Fig.~\ref{fig:injapps}c,d). These are single runs for each start and method, using CPU float64, 1024 particles, a 1~ps step and the cathode image. The three methods run concurrently in separate processes, with five threads each on the 16-core CPU; elapsed times include contention and each method's backward pass or surrogate fit.

Bayesian optimisation finishes with the lowest objective from both starts, 0.381 and 0.382, and comes within 10\% of the best objective found from that start in 291 and 494~s. AD uses the fewest evaluations to reach that threshold: 4 and 10, compared with 13 and 22 for Bayesian optimisation and 22 and 33 for Nelder--Mead. AD then stops improving, at 0.392 and 0.403, with the gun phase at $+1.5^\circ$ and $-2.1^\circ$; the best runs of the other two methods have it within $0.1^\circ$ of the $+8^\circ$ bound of the search range, and Nelder--Mead finishes at 0.384 and 0.385.

An AD evaluation costs about six forward runs. At five parameters, its smaller evaluation count does not compensate for that cost in elapsed time: AD takes 529 and 1322~s to come within 10\%, against 291 and 494~s for Bayesian optimisation. This contrasts with 24-parameter ring matching, where the larger FD gradient cost makes AD advantageous. The comparison describes these starts and the stated concurrent resource allocation.

\subsection{Space-charge-dominated drift}
\label{sec:drift}

\begin{figure*}[t]
\includegraphics[width=\textwidth]{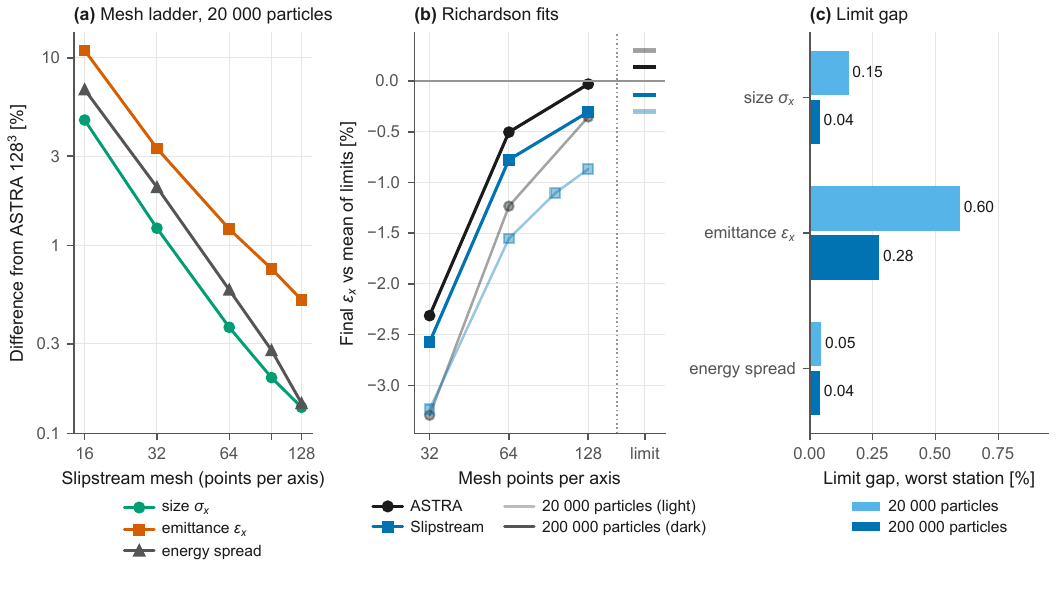}
\caption{Mesh convergence of the 20~pC simulation through a 3~m drift. (a) Difference of \textsc{Slipstream}'s Cartesian results from \textsc{ASTRA} at its finest ($128^3$) mesh against \textsc{Slipstream} mesh, maximum over stations, 20\,000 particles. (b) Final-station emittance of each solver against the mean of the two extrapolated limits, for 20\,000 (light) and 200\,000 (dark) particles; ticks at ``limit'' mark each solver's extrapolated value. (c) Difference between the two extrapolated limits, worst over stations.}
\label{fig:convergence}
\end{figure*}

\paragraph*{Cross-code agreement.} Simulating a drift separates space-charge effects from magnetic transport. Against \textsc{ASTRA}'s $32^3$ mesh, all tested settings of \textsc{Slipstream}, \textsc{Cheetah}, \textsc{Ocelot}, \textsc{GPT}, \textsc{ImpactX} and \textsc{ASTRA} itself meet the size, length and emittance tolerances. Energy spread is more discriminating: differences range from 0.2\% to 9\%. \textsc{Slipstream}'s Cartesian solver differs by 0.06\% in size, 0.07\% in emittance and 0.9\% in spread. Its GPU float32 and CPU float64 results agree to $3\times10^{-6}$.

\begin{figure*}[t]
\includegraphics[width=\textwidth]{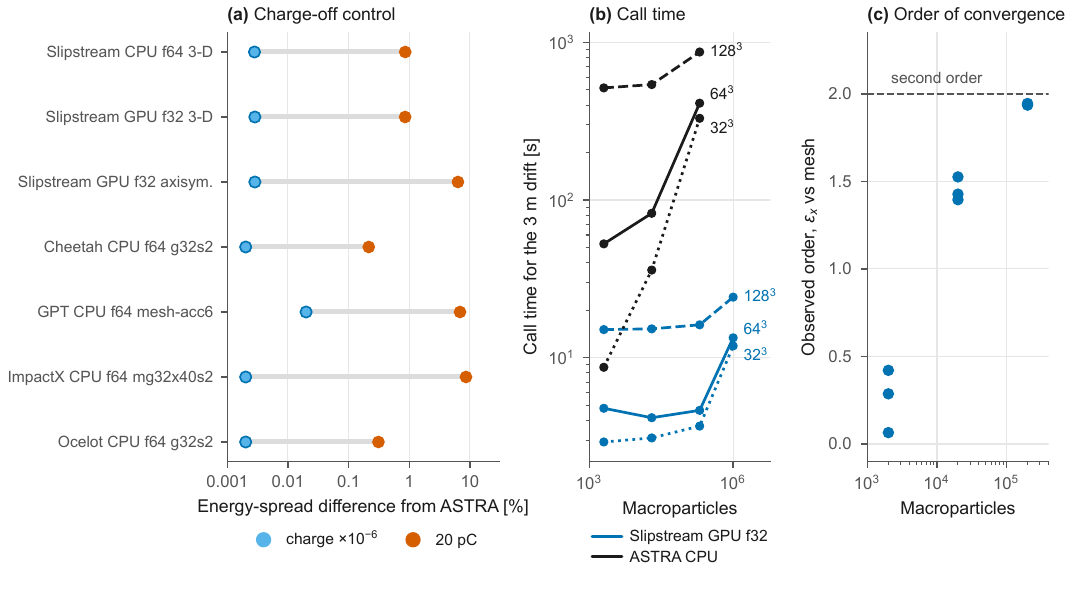}
\caption{Charge-off control and particle-count ladder for the drift. (a) Energy-spread difference from \textsc{ASTRA} at negligible charge (blue) and at 20~pC (orange), one row per setting. (b) Call time against macroparticle number for \textsc{Slipstream}'s GPU Cartesian presets and \textsc{ASTRA}'s three-dimensional solver (single runs for the slowest points). (c) Observed order of convergence of the emittance with mesh, for three independent bunches at each particle count.}
\label{fig:drift}
\end{figure*}

Reducing the charge by $10^{-6}$ reduces all differences to 0.02\% or below [Fig.~\ref{fig:drift}(a)], attributing the larger charged-bunch differences to the space-charge calculations. Space charge changes the energy spread by 17\% in the three-dimensional solvers and 10--12\% in \textsc{GPT}, \textsc{ImpactX} and \textsc{Slipstream}'s axisymmetric solver. \textsc{Slipstream}'s Cartesian discrepancy is about 5\% of the space-charge-induced change, illustrating why agreement should also be judged relative to the physical effect being calculated.

The mean energy gain over the drift is 30.5~eV in \textsc{Slipstream} and 31.0~eV in \textsc{ASTRA} at their finest meshes. A Gaussian-bunch field-energy estimate gives 27.5~eV, providing an additional scale for this small energy transfer.

\begin{table}[t]
\caption{\textsc{Slipstream} Cartesian mesh ladder for the 20\,000-particle drift (1~ps step, field refresh every 10 steps). Errors are maxima over stations against \textsc{ASTRA} at its finest mesh ($128^3$). Chg. is the change from the next coarser mesh.}
\label{tab:mesh}
\footnotesize
\setlength{\tabcolsep}{2pt}
\begin{tabular}{@{}rrrrrrr@{}}
\toprule
Mesh & Warm & $\epsilon_x$ & Spread & Size & $\epsilon_x$ & Spread \\
 & [s] & chg. & chg. & \multicolumn{3}{c}{vs \textsc{ASTRA} $128^3$} \\
\midrule
$16^3$ & 3.09 & --- & --- & 4.7\% & 11\% & 6.8\% \\
$32^3$ & 3.09 & 8.6\% & 5.1\% & 1.2\% & 3.3\% & 2.0\% \\
$64^3$ & 4.16 & 2.2\% & 1.5\% & 0.37\% & 1.2\% & 0.58\% \\
$96^3$ & 14.9 & 0.54\% & 0.31\% & 0.20\% & 0.75\% & 0.28\% \\
$128^3$ & 15.2 & 0.27\% & 0.13\% & 0.14\% & 0.51\% & 0.14\% \\
\bottomrule
\end{tabular}
\end{table}

\paragraph*{Mesh refinement.} \textsc{ASTRA} rounds its three-dimensional mesh to powers of two, giving effective meshes of $32^3$, $64^3$ and $128^3$. Its emittance and energy spread change by 0.9\% and 0.2\%, respectively, between the finest two. \textsc{Slipstream}'s difference from \textsc{ASTRA}'s finest mesh decreases with refinement (Table~\ref{tab:mesh}, Fig.~\ref{fig:convergence}a). Its $64^3$ mesh takes about 4~s; halving either the particle-advance step or the field-refresh interval changes its results by less than 0.01\%.

To estimate zero-grid-spacing limits, each solver's mesh series is fitted at each station to $f(N)=f_\infty+CN^{-p}$, with $N$ the number of cells per dimension and $p$ fitted from the data~\cite{roache1997}. At 200\,000 particles, the fitted size, energy-spread and emittance results approach second-order convergence, with small differences in  extrapolated limits [Fig.~\ref{fig:convergence}(b,c)], whilst the $128^3$ results are within 0.2\% of their respective fitted limits.

At 20\,000 particles, the fitted emittance order is 1.4--1.5 and the limits differ by $<1$\%. The small residuals at higher particle counts have not been uniquely attributed; field smoothing and boundary treatment are possible contributors.

\paragraph*{Particle number and cost.} The scaling study draws bunches with 2000 to $10^6$ macroparticles [Fig.~\ref{fig:drift}(b,c)]. \textsc{Slipstream}'s $64^3$ GPU calculation takes 4--5~s up to 200\,000 particles and 13~s at $10^6$, showing that the mesh dominates its cost over much of this range. \textsc{ASTRA} at $32^3$ takes 9, 36 and 330~s for 2000, 20\,000 and 200\,000 particles. Comparing both codes at $32^3$, \textsc{Slipstream} on the GPU is 3 times faster at 2000 particles and 90 times faster at 200\,000.

Particle count also affects mesh convergence. The observed emittance order-of-convergence rises from below 0.5 at 2000 particles to 1.4--1.5 at 20\,000 and 1.94 at 200\,000, consistently for three independent bunches. It reaches 2.0 at $10^6$ particles. Thus, finer grids alone are insufficient when too few particles sample each cell. At matched meshes, the \textsc{Slipstream}--\textsc{ASTRA} emittance difference settles near 0.3\%.

\subsection{Field-map injector: CLARA gun and first accelerating structure}
\label{sec:clara}

\begin{figure*}[t]
\includegraphics[width=\textwidth]{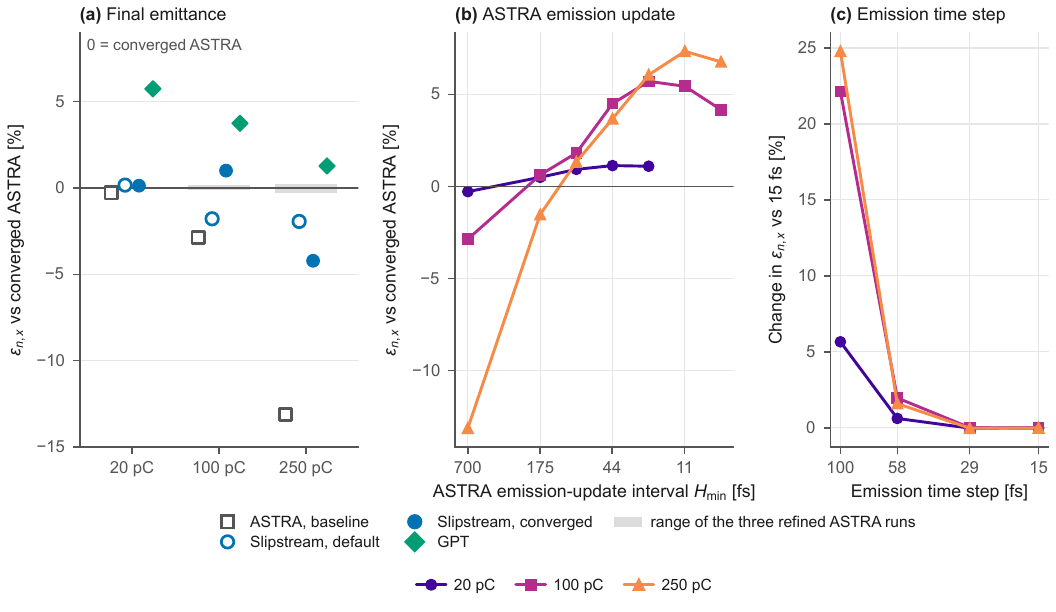}
\caption{CLARA injector. (a) Final normalised horizontal emittance relative to converged \textsc{ASTRA} at three charges: \textsc{ASTRA} at its baseline settings (open squares), \textsc{Slipstream} at its default (open circles) and converged settings (filled circles), and \textsc{GPT} (diamonds); the shaded band spans the minimum to the maximum of the three refined \textsc{ASTRA} settings. (b) \textsc{ASTRA}'s emittance against the interval at which the space-charge field is refreshed during emission, with the baseline field-rescaling limit and mesh, relative to converged \textsc{ASTRA}. (c) Sensitivity of \textsc{Slipstream}'s emittance to the emission time step, relative to its 15~fs run.}
\label{fig:clara}
\end{figure*}

The CLARA comparison in this section asks how numerical settings affect agreement through a gun and accelerating structure. It tracks a 1.5-cell gun at 120~MV/m with its solenoid, a 60-cell travelling-wave structure at 21~MV/m and two linac solenoids. The observation plane is at about 35~MeV. Charges examined are 20, 100 and 250~pC.

The codes use the same on-axis field tables and $262\,144$ macroparticles. \textsc{Slipstream} uses float32 on the GPU, its axisymmetric solver and a cathode image. \textsc{GPT} uses a mesh solver with about $64^3$ lines through the bunch and a conducting cathode plane. Its travelling-wave structure is represented by two standing-wave maps in quadrature; cavity phases are set from the single-particle crest plus \textsc{ASTRA}'s relative phase.

\textsc{ASTRA} and \textsc{Slipstream} are each compared at a faster setting and after numerical refinement to improve convergence (Table~\ref{tab:clara}). The faster settings are the baseline CLARA \textsc{ASTRA} deck and \textsc{Slipstream}'s package defaults. Rows labelled ``converged'' identify the refined settings tested here, rather than an exact physical solution. \textsc{ASTRA} supplies the common numerical baseline. \textsc{GPT} uses its default mesh at all three charges; separate particle and mesh checks are available at 20~pC.

\begin{table*}[t]
\caption{Final normalised horizontal emittance $\epsilon_{n,x}$ [mm\,mrad] at $z=3.37147$~m for refined \textsc{ASTRA} and \textsc{Slipstream}, default-mesh \textsc{GPT}, and the faster \textsc{ASTRA} baseline and \textsc{Slipstream} defaults. \textsc{GPT} refinement checks are restricted to 20~pC. The next columns give the difference of each run from converged \textsc{ASTRA} in percent (positive: above \textsc{ASTRA}). Run times: \textsc{ASTRA} on 16 CPU threads, \textsc{Slipstream}'s tracking loop on the GPU, \textsc{GPT} on 12 CPU threads.}
\label{tab:clara}
\footnotesize
\setlength{\tabcolsep}{4pt}
\begin{tabular}{@{}llrrrrrrr@{}}
\toprule
Charge & Code & $\epsilon_{n,x}$ & \multicolumn{5}{c}{Difference from converged \textsc{ASTRA} [\%]} & Run \\
\cmidrule(lr){4-8}
 & & [mm\,mrad] & $\epsilon_{n,x}$ & $\sigma_x$ & $\sigma_z$ & $\sigma_E$ & $\epsilon_{n,z}$ & time \\
\midrule
20~pC & \textsc{ASTRA}, converged & 0.3117 & -- & -- & -- & -- & -- & 89.1~min \\
 & \textsc{Slipstream}, converged & 0.3121 & $+0.1$ & $-4.5$ & $-0.8$ & $+1.9$ & $+1.7$ & 82~s \\
 & \textsc{GPT} & 0.3296 & $+5.7$ & $+2.9$ & $-2.9$ & $+3.4$ & $+5.3$ & 27.5~min \\ \addlinespace[2pt]
 & \textsc{ASTRA}, baseline & 0.3108 & $-0.3$ & $-0.1$ & $+0.2$ & $+0.6$ & $+0.6$ & 8.7~min \\
 & \textsc{Slipstream}, default & 0.3122 & $+0.2$ & $-4.1$ & $-0.7$ & $+2.1$ & $+2.1$ & 14~s \\
\midrule
100~pC & \textsc{ASTRA}, converged & 0.4274 & -- & -- & -- & -- & -- & 91.3~min \\
 & \textsc{Slipstream}, converged & 0.4316 & $+1.0$ & $-5.5$ & $-1.0$ & $+2.7$ & $+0.6$ & 4.2~min \\
 & \textsc{GPT} & 0.4434 & $+3.7$ & $+5.6$ & $-3.2$ & $+4.9$ & $+1.2$ & 36.2~min \\ \addlinespace[2pt]
 & \textsc{ASTRA}, baseline & 0.4151 & $-2.9$ & $+0.2$ & $+0.5$ & $+1.3$ & $+4.8$ & 9.2~min \\
 & \textsc{Slipstream}, default & 0.4197 & $-1.8$ & $-5.1$ & $-0.9$ & $+3.0$ & $+0.9$ & 17~s \\
\midrule
250~pC & \textsc{ASTRA}, converged & 0.5560 & -- & -- & -- & -- & -- & 111~min \\
 & \textsc{Slipstream}, converged & 0.5325 & $-4.2$ & $-3.7$ & $-1.9$ & $+3.4$ & $-1.6$ & 2.2~min \\
 & \textsc{GPT} & 0.5631 & $+1.3$ & $+2.3$ & $-4.8$ & $+7.9$ & $-0.1$ & 32.1~min \\ \addlinespace[2pt]
 & \textsc{ASTRA}, baseline & 0.4830 & $-13.1$ & $+0.1$ & $+0.8$ & $+2.8$ & $+13.0$ & 10.5~min \\
 & \textsc{Slipstream}, default & 0.5452 & $-1.9$ & $-4.1$ & $-1.6$ & $+4.3$ & $-0.8$ & 22~s \\
\bottomrule
\end{tabular}
\end{table*}

\paragraph*{Agreement after refinement.} Final horizontal emittances from all three codes are within 6\% of the refined \textsc{ASTRA} result at each charge (Table~\ref{tab:clara}, Fig.~\ref{fig:clara}a). \textsc{Slipstream} differs by $+0.1\%$, $+1.0\%$ and $-4.2\%$ at 20, 100 and 250~pC. \textsc{GPT} is 1--6\% above \textsc{ASTRA}, with a smaller excess at higher charge.

Agreement in emittance does not imply agreement in every observable. \textsc{Slipstream} gives smaller rms beam sizes than \textsc{ASTRA}, while \textsc{GPT} gives larger sizes. Both predict slightly shorter bunches and higher energy spreads, with larger deviations in \textsc{GPT}. Final mean energies agree to 0.3\%. No solver is assumed to give the exactly correct physical solution.

At 20~pC, doubling \textsc{GPT}'s mesh resolution changes the emittance by 0.2\% and increasing the particle number fourfold from $65\,536$ changes it by 1.8\%, so the 5.7\% difference from \textsc{ASTRA} is not a resolution effect of either code; its cause is not identified, and convergence at the other charges was not tested. The $-4.2\%$ difference of refined \textsc{Slipstream} at 250~pC has not been traced to a specific cause either; its default settings differ by $-1.9\%$, so the numerical settings move the result by about two percentage points. The three codes nevertheless agree to within 6\% on the emittance, and converged \textsc{Slipstream} and \textsc{ASTRA} agree to 1\% at 20 and 100~pC.

\begin{figure*}[t]
\includegraphics[width=\textwidth]{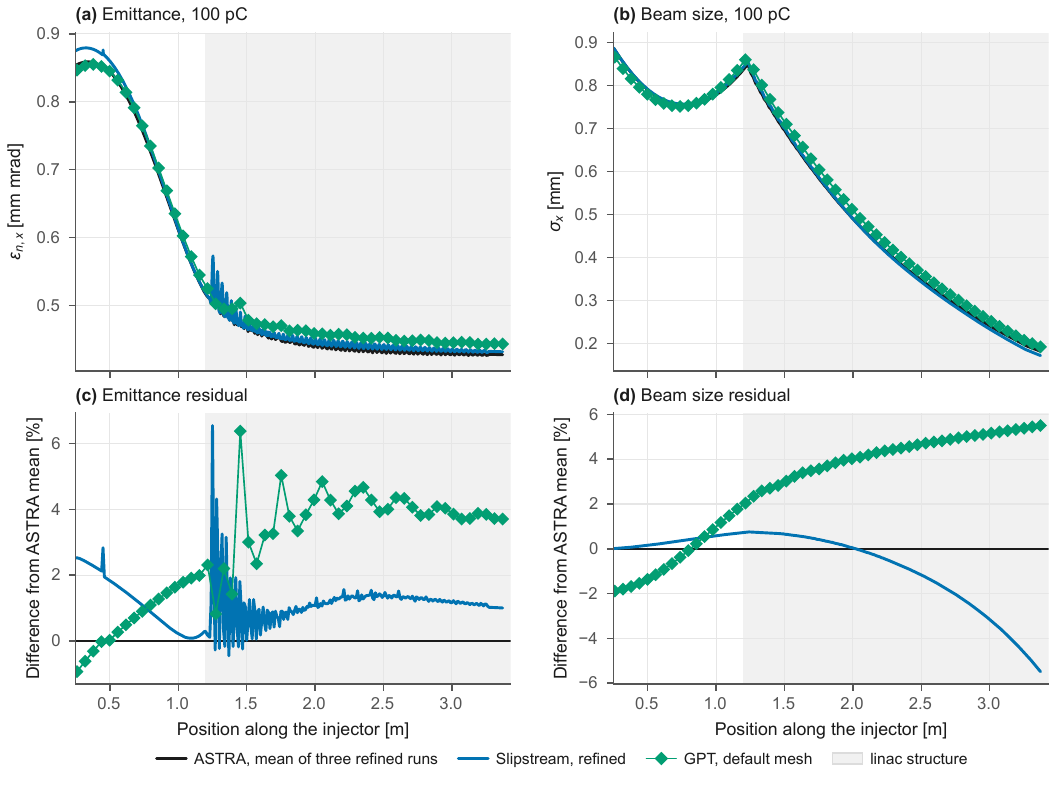}
\caption{Beam evolution in the 100~pC CLARA injector. (a, b) Normalised horizontal emittance and rms horizontal size for the mean of the three refined \textsc{ASTRA} runs (which differ by less than 0.4\%, too little to draw as a band), refined \textsc{Slipstream}, and \textsc{GPT} with its default mesh and $262\,144$ macroparticles, with snapshots every 0.2~ns. Emittances are canonical. (c, d) Differences from the \textsc{ASTRA} mean, evaluated for \textsc{Slipstream} at \textsc{ASTRA}'s output positions (every 10~mm) and for \textsc{GPT} at its snapshot positions. The shaded region marks the linac structure. The output spacing is 2~mm for \textsc{Slipstream}, 10~mm for \textsc{ASTRA} and 6~cm for \textsc{GPT}.}
\label{fig:claraprofile}
\end{figure*}

\paragraph*{Evolution along the line.} Figure~\ref{fig:claraprofile} follows the 100~pC emittance and beam size from the gun to the observation plane for \textsc{ASTRA} (the mean of the three refined runs), refined \textsc{Slipstream} and \textsc{GPT} with its default mesh. Up to the accelerating structure the codes match each other: \textsc{Slipstream}'s emittance is 0.1--2.8\% above the \textsc{ASTRA} mean and \textsc{GPT}'s is between $-0.9$ and $+2.0$\%, and the beam sizes differ by less than 0.8\% for \textsc{Slipstream} and 2\% for \textsc{GPT}. The differences grow in and after the accelerating structure. Over its first 0.4~m \textsc{Slipstream}'s emittance oscillates with periods of 1 to 3~cm, and its difference from \textsc{ASTRA} ranges from $-0.5$ to $+6.6$\%. The output spacing is 2~mm for \textsc{Slipstream}, 10~mm for \textsc{ASTRA} and 6~cm for \textsc{GPT}, so the size of this oscillation, and of the difference there, depends on the sampling. Beyond 1.6~m \textsc{Slipstream}'s emittance is 0.2--1.6\% above \textsc{ASTRA}'s and \textsc{GPT}'s is 3.2--5.0\% above, and the beam sizes move in opposite directions, ending $-5.5$\% and $+5.5$\% from \textsc{ASTRA}'s.

\paragraph*{Accuracy and runtime.} \textsc{ASTRA}'s CLARA baseline settings run 10--11 times faster than its refined settings. The emittance differs by less than 3\% at 20 and 100~pC, but is 13\% lower at 250~pC. The emission-field refresh is important: the baseline recalculates this field only about four times over the 2.92~ps emission window. Shortening this interval raises the calculated emittance, which levels off around 22~fs after the emission [Fig.~\ref{fig:clara}(b)].

\textsc{Slipstream}'s defaults run 6--15 times faster than its refined settings. Default-setting emittances lie within 2\% of refined \textsc{ASTRA} at every charge. At 250~pC the default is closer to \textsc{ASTRA} than \textsc{Slipstream}'s own refined result. That closeness alone does not establish a more accurate or better-converged prediction: numerical changes can partly compensate for differences between solvers.

\paragraph*{Numerical choices in \textsc{Slipstream}.} Three choices materially affect the result. The emission step must resolve the 2.92~ps window [Fig.~\ref{fig:clara}(c)]. Omitting the cathode image raises emittance by up to 60\%. At 100~pC, refining the radial grid from the default 64 cells to 256 raises emittance by 4\%. These sensitivities explain why particle number, mesh and emission controls must accompany any quoted agreement.

\paragraph*{Structure wake.} The travelling-wave structure has one monopole-wake kick at the centre of each of its 60 cells (Sec.~\ref{sec:injector}). At 250~pC, using baseline settings in both codes, the wake lowers the final mean energy by about 40~keV from 35~MeV and reduces the energy spread by 12--13\%, with no change in transverse emittance.

\textsc{Slipstream} reproduces \textsc{ASTRA}'s wake-induced changes in energy and energy spread to 6\%, and a single kick's energy loss to 1\%. Using \textsc{ASTRA}'s own charge profile, the \textsc{Slipstream} kernel reproduces its wake potential to 0.01\% of the peak. We did not isolate how much of the 6\% difference comes from the different incoming bunch distributions. Adding the wake produces no measurable increase in tracking-loop time.

\paragraph*{Runtime.} Table~\ref{tab:clara} keeps the accuracy settings and their costs separated. \textsc{Slipstream}'s default tracking-loop times are 14, 17 and 22~s for $262\,144$ macroparticles on the RTX 3050. Its refined settings take 82~s, 4.2~min and 2.2~min, respectively.

At the baseline/default settings, \textsc{Slipstream}'s GPU tracking loop is 29--38 times faster than \textsc{ASTRA} 4.0 on 16 CPU threads and 90--130 times faster than \textsc{GPT} on 12 threads. \textsc{ASTRA} uses Open MPI in the \textsc{simcodes} container under WSL2, with ranks on hardware threads. It reproduces the serial emittance; the serial calculation takes 63--67~min. These timings compare the stated implementations and resource allocations, including \textsc{Slipstream}'s tracking loop rather than its initialisation.

\section{Additional validation cases}
\label{sec:further}

\subsection{Coherent synchrotron radiation}
\label{sec:csr}

The CSR study here separates tests of the individual wake models from transport through a full bunch compressor. The references for these wake models use the same one-dimensional line-charge approximation but are calculated independently of the tracking implementation: a closed-form Gaussian-bunch loss and a direct evaluation of retarded Li\'enard--Wiechert fields~\cite{jackson1999} along straight sections and circular arcs. Agreement with these references tests numerical implementation within that approximation.

\begin{figure*}[t]
\includegraphics[width=\textwidth]{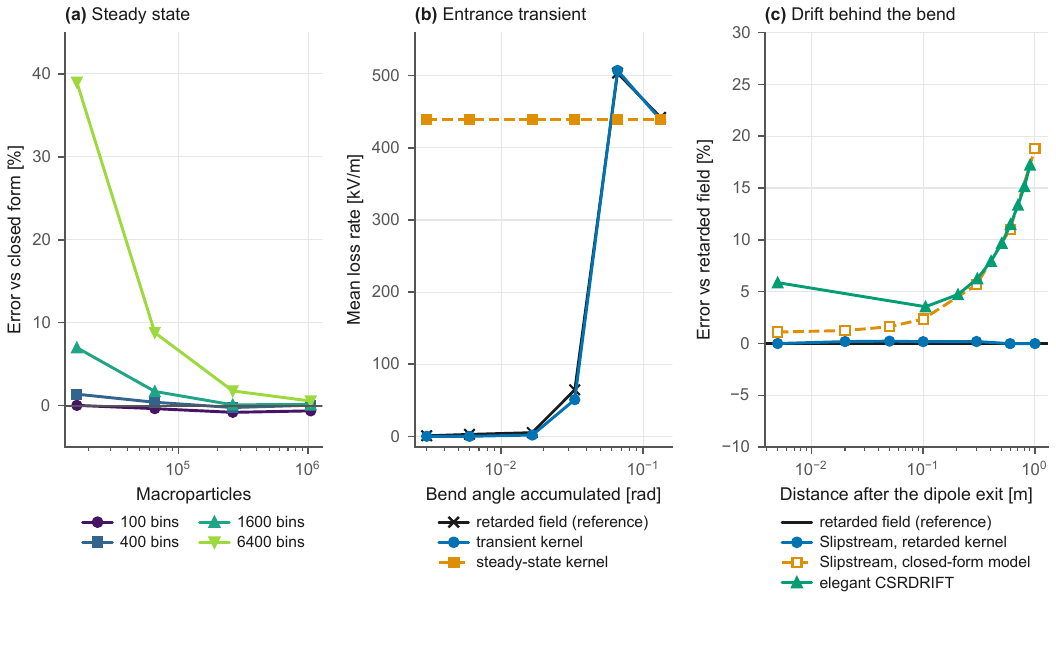}
\caption{CSR wake models against closed-form and retarded-field references for a 20~$\mu$m, 100~pC Gaussian bunch at 190~MeV on the geometry of a CLARA chicane dipole. (a) Error of the bunch-mean energy loss of the steady-state wake against the closed form calculation, versus particle number for four bin counts. (b) Bunch-mean loss rate of the retarded-field reference, the entrance-transient wake model and the steady-state wake model, against the bend angle accumulated inside the dipole; the transient wake model gives zero loss at the two smallest angles. (c) Error of the mean loss rate behind the dipole exit against the retarded field: \textsc{Slipstream}'s retarded-field wake model and closed-form model, and the \texttt{CSRDRIFT} wake of \textsc{elegant}.}
\label{fig:csr}
\end{figure*}

\textsc{Slipstream} evaluates the wake on a fixed histogram grid. Inside a dipole, the entrance transient describes the build-up towards steady state over about 0.1~m for this bunch and geometry~\cite{brynes2018}. Behind the dipole, a tabulated retarded-field wake model describes the drift wake. The steady-state wake model agrees with the closed-form loss to 0.2\%. The entrance-transient mean loss agrees with the direct retarded-field calculation to 1\% at the last two sampled angles, 0.07 and 0.13~rad; larger differences occur earlier, where the loss is small [Fig.~\ref{fig:csr}(b)]. The retarded-drift mean loss agrees to 1\% over the sampled downstream points [Fig.~\ref{fig:csr}(c)]. \textsc{Elegant}'s closed-form drift model~\cite{stupakov2002} gives up to 17\% greater loss in the tested drift, with the difference increasing downstream. \textsc{Slipstream}'s own closed-form version of this model shows the same trend: it is 1--2\% high up to 5~cm from the exit, 11\% high at 0.6~m and 19\% high at 1~m.

\begin{figure*}[t]
\includegraphics[width=\textwidth]{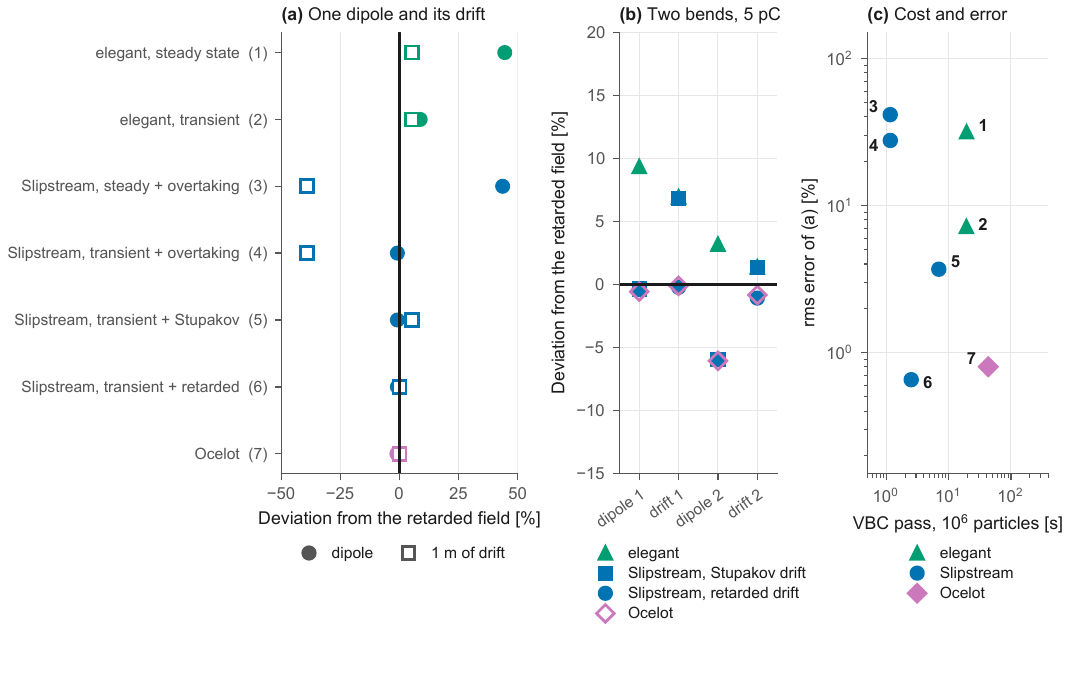}
\caption{CSR models against the retarded-field reference. (a) Mean energy loss of a 20~$\mu$m, 100~pC Gaussian bunch in one VBC dipole (circles) and the following 1~m of drift (squares), as a deviation from the retarded field, for \textsc{elegant} (steady-state and transient models, closed-form drift model), \textsc{Slipstream} (steady-state or transient model with the overtaking-length, closed-form or retarded drift model) and \textsc{Ocelot}. (b) Deviation of the loss in each segment of the first two VBC dipoles at 5~pC from the retarded field of the frozen bunch on the two-bend path. (c) Run time of a VBC pass with $10^6$ particles (Table~\ref{tab:csrtime}; \textsc{Slipstream} on the GPU in double precision, \textsc{elegant} and \textsc{Ocelot} on the CPU) against the rms error of the two values in (a); the numbers refer to the settings listed in (a).}
\label{fig:csrmodels}
\end{figure*}

Integrating the loss over one dipole and its following drift shows the effect of each approximation (Fig.~\ref{fig:csrmodels}a). \textsc{Slipstream}'s transient and retarded-drift wake models reproduce the reference loss in each segment to 1\%, as does \textsc{Ocelot}'s retarded-field CSR process~\cite{agapov2014}. A steady-state model throughout the dipole overestimates the loss by 44\% because it omits the build-up. A steady-state drift truncated at the overtaking length underestimates drift loss by 39\%. With 8 dipole slices, \textsc{elegant}'s transient model is 9\% higher and requires 64 slices for convergence in this test; \textsc{Slipstream} is converged at 8 slices.

A two-bend test uses the first two dipoles of the CLARA~\cite{angalkalinin2020} variable bunch compressor (VBC), at 5~pC, where bunch shape changes little (Fig.~\ref{fig:csrmodels}b). \textsc{Slipstream} and \textsc{Ocelot} agree with the retarded-field loss of the frozen bunch to 2\% in total, but both underestimate the second-dipole loss by 6\%. The direct reference includes radiation carried from the first bend, which these tested models omit.

\begin{table*}[t]
\caption{Mean CSR energy loss and normalised horizontal emittance of a 100~pC bunch through the CLARA bunch compressor, for three chicane strengths ($0.925$ is full compression), in \textsc{elegant}, \textsc{Ocelot} and \textsc{Slipstream}. The emittance without CSR is 0.483~$\mu$m. ``Dipoles only'' has no drift CSR; ``dipoles and drifts'' uses \textsc{elegant}'s transient + drift model, \textsc{Ocelot}'s retarded-field process, and \textsc{Slipstream}'s entrance transient with retarded-field drift kernel.}
\label{tab:csr}
\small
\begin{tabular}{llcccccc}
\toprule
 & & \multicolumn{2}{c}{0.8} & \multicolumn{2}{c}{0.925} & \multicolumn{2}{c}{1.0} \\
 & & loss [keV] & $\epsilon_{n,x}$ [$\mu$m] & loss [keV] & $\epsilon_{n,x}$ [$\mu$m] & loss [keV] & $\epsilon_{n,x}$ [$\mu$m] \\
\midrule
Dipoles only & \textsc{elegant} & 9.9 & 0.4995 & 165.3 & 0.9416 & 65.9 & 0.8030 \\
 & \textsc{Slipstream} & 8.5 & 0.4950 & 153.7 & 0.8858 & 61.0 & 0.7659 \\
Dipoles and drifts & \textsc{elegant} & 44.4 & 0.5548 & 373.7 & 1.7442 & 126.7 & 1.1156 \\
 & \textsc{Ocelot} & 43.3 & 0.5417 & 356.0 & 1.6168 & 126.5 & 1.1162 \\
 & \textsc{Slipstream} & 44.1 & 0.5442 & 332.8 & 1.6036 & 115.2 & 1.0283 \\
\bottomrule
\end{tabular}
\end{table*}

The full-compressor calculation tests an evolving bunch rather than an isolated wake model (Table~\ref{tab:csr}). All three codes give the same emittance without CSR. With CSR only in the dipoles, \textsc{Slipstream}'s energy loss is 7--14\% below \textsc{elegant}'s, consistent with sensitivity to the transient slicing.

Including the drifts brings the energy losses within 3\% at the weakest chicane strength. At nominal strength, \textsc{Ocelot} and \textsc{elegant} give nearly the same loss, while \textsc{Slipstream} is 9\% lower. At full compression, the losses are 332.8~keV in \textsc{Slipstream}, 356.0~keV in \textsc{Ocelot} and 373.7~keV in \textsc{elegant}: the \textsc{Slipstream}--\textsc{elegant} difference is about 11\%. The emittances also differ, as shown in the table. The 1\% isolated model agreement therefore does not extend to every observable.

\begin{table}[t]
\caption{Run time [s] of one VBC pass with CSR for $10^6$ and $10^5$ particles (400 bins), with the speed-up over \textsc{elegant}'s transient model in parentheses. \textsc{Elegant} is serial and includes its file input and output; \textsc{Ocelot} (version 26.6.1, Numba) runs on the CPU in WSL2; \textsc{Slipstream} (retarded drift kernel) runs on the CPU with one and 32 threads and on the GPU (RTX 3050). The lower block gives \textsc{Slipstream} on the GPU in double precision for each CSR model.}
\label{tab:csrtime}
\small
\begin{tabular}{lcc}
\toprule
 & $10^6$ particles & $10^5$ particles \\
\midrule
\textsc{elegant}, serial CPU & 19.3 (1.0) & 1.76 (1.0) \\
\textsc{Ocelot}, CPU & 43.3 (0.45) & 6.1 (0.29) \\
\textsc{Slipstream}, one CPU thread & 25.1 (0.77) & 1.42 (1.2) \\
\textsc{Slipstream}, 32 CPU threads & 8.65 (2.2) & 0.84 (2.1) \\
\textsc{Slipstream}, GPU single & 1.61 (12) & 0.54 (3.2) \\
\textsc{Slipstream}, GPU double & 2.51 (7.7) & 0.59 (3.0) \\
\midrule
\multicolumn{3}{l}{\emph{\textsc{Slipstream}, GPU double, by CSR model}} \\
steady state & 1.16 (17) & 0.15 (12) \\
transient & 1.16 (17) & 0.18 (9.8) \\
transient, closed-form drift & 6.96 (2.8) & 1.57 (1.1) \\
transient, retarded drift & 2.51 (7.7) & 0.59 (3.0) \\
\bottomrule
\end{tabular}
\end{table}

For $10^6$ particles, the double-precision GPU calculation with entrance transient and retarded-drift kernel takes 2.51~s, 7.7 times faster than serial \textsc{elegant} (Table~\ref{tab:csrtime}, Fig.~\ref{fig:csrmodels}c). \textsc{Slipstream} with 32 CPU threads is 2.2 times faster than \textsc{elegant}; \textsc{Ocelot} takes about twice as long. \textsc{Elegant}'s times include file input and output, so the ratios describe these workflows and resource allocations.

The drift treatment has a substantial cost. The closed-form model is evaluated at centimetre intervals, as in \textsc{elegant}. The retarded-drift version takes about one third as long, using a fused \textsc{Warp} kernel that is 30 times faster than the \textsc{PyTorch} bisection calculation it replaces. Direct pairwise quadrature takes minutes per pass at $10^5$ particles in double precision on this GPU, whose double-precision arithmetic rate is much lower than its single-precision rate.

\subsection{Booster dynamics, scattering and impedance}\label{subsec:booster}

\paragraph*{Booster dynamics.} The Diamond-II booster~\cite{diamond2019} provides a larger-ring validation test, and includes combined-function dipole-quadrupole-sextupole magnets. \textsc{Slipstream} agrees with \textsc{elegant}'s Twiss tunes to $5\times10^{-5}$ and chromaticities to 0.02\%, and reproduces their dependence on momentum offset. A dynamic-aperture test tracks an initial-position grid for 1000 turns at three momentum offsets. The stable areas agree with \textsc{elegant} to 3\% in GPU float32; launches classified differently lie near the unstable boundary.

Frequency-map diffusion is more sensitive to precision. Float32 adds about three orders of magnitude of noise relative to double-single arithmetic, which is therefore used for the higher-precision maps. The GPU dynamic-aperture grid takes 9~s, 75 times faster than serial \textsc{elegant} and 11 times faster than \textsc{Pelegant}, the MPI-parallel version of \textsc{elegant}, with 16 ranks. Against \textsc{Pelegant} on 32 ranks, the float32 frequency map is 7.4 times faster, but double-single arithmetic is 2.3 times slower. The precision required by the observable can drastically change any performance conclusion.

\paragraph*{Intrabeam scattering.} In another benchmark, Bjorken--Mtingwa rates~\cite{bjorken1983} averaged element by element around the ELENA lattice~\cite{chohan2014} agree with \textsc{Xsuite}'s \texttt{BjorkenMtingwaIBS} calculation to within 5\% in all three planes for the same lattice and beam. Evaluating the rates once at the average optics instead differs by 12--19\%. This comparison tests the lattice averaging as well as the growth-rate calculation.

\paragraph*{Resistive-wall impedance.} In a final benchmark model, \textsc{Slipstream} solves a round chamber impedance with multiple resistive layers by matching fields across their interfaces. A Yokoya form factor supplies an approximate longitudinal impedance for a flat chamber~\cite{yokoya1993}. The test chamber is the non-evaporable-getter (NEG) coated HALF vacuum chamber~\cite{bai2021}: 13~mm radius, a 1~$\mu$m NEG layer of resistivity $10^{-5}~\Omega$\,m on a copper-like substrate, and a 2.2~GeV electron beam.

\begin{figure*}[t]
\includegraphics[width=\textwidth]{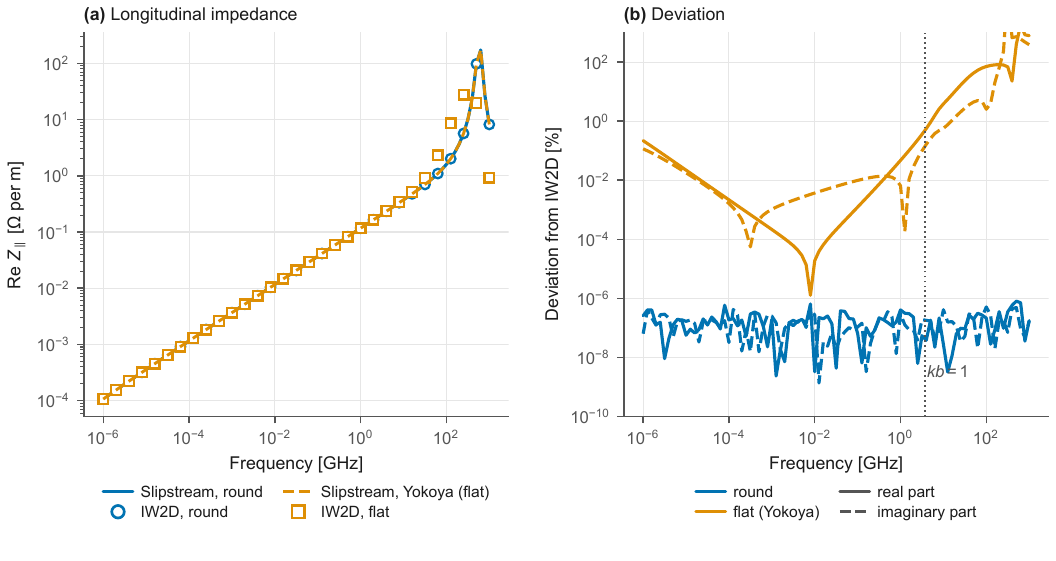}
\caption{Resistive-wall impedance of the NEG-coated HALF main chamber (13~mm radius or half gap, 1~$\mu$m NEG, 2.2~GeV, 1~m long). (a) Real part of the longitudinal impedance: \textsc{Slipstream}'s round-chamber solution and \textsc{IW2D} for the round chamber, and \textsc{IW2D}'s flat-chamber solver and \textsc{Slipstream}'s round result with Yokoya's factor for parallel plates. (b) Deviation of \textsc{Slipstream}'s values from \textsc{IW2D}; the dotted line marks $kb=1$.}
\label{fig:imp}
\end{figure*}

For the round chamber, longitudinal and dipolar impedances agree with an \textsc{IW2D} run~\cite{mounet2010} to a maximum relative difference of $3\times10^{-8}$ over 1~kHz--1~THz (Fig.~\ref{fig:imp}b). \textsc{Slipstream} takes 7~ms, compared with 0.4~s for \textsc{IW2D} including start-up. The close numerical agreement checks the layer-matching implementation; this timing comparison includes the executable start-up cost.

For parallel plates, \textsc{IW2D}'s flat-chamber longitudinal result agrees with the round result and Yokoya factor 1 to within 1\% up to 5~GHz. The two depart as $kb$ approaches unity, at about 3.7~GHz, but the difference stays below 1\% until 5~GHz and reaches 4\% at 10~GHz. Here $k$ is the wave number and $b$ the half gap. This sets a frequency-dependent limit on the flat-chamber approximation. \textsc{Slipstream} currently only provides longitudinal impedance for flat chambers, so \textsc{IW2D}'s transverse flat-chamber outputs are not compared.

\section{Discussion}
\label{sec:discussion}

\begin{table*}[t]
\caption{Selected model and execution differences relevant to these code benchmarks, from the cited papers, project pages and source trees. Only the differentiable or GPU codes are listed; \textsc{elegant}, \textsc{ASTRA}, \textsc{GPT} and \textsc{Ocelot} are omitted. ``n.s.'' means not stated in the sources consulted. The last column lists the benchmark cases of Sec.~\ref{sec:results} each code took part in; versions are in Table~\ref{tab:versions}.}
\label{tab:codes}
\footnotesize
\setlength{\tabcolsep}{3pt}
\renewcommand{\arraystretch}{1.1}
\begin{tabular}{@{}llllll@{}}
\toprule
\pcell{1.7cm}{Code} & \pcell{2.7cm}{Execution} & \pcell{3.3cm}{Differentiability} & \pcell{3.1cm}{Nonlinear magnets} & \pcell{3.2cm}{Space charge} & \pcell{2.5cm}{Benchmarked} \\
\midrule
\pcell{1.7cm}{\textsc{Slipstream}} & \pcell{2.7cm}{\textsc{PyTorch}, optional \textsc{Warp}; CPU, CUDA} & \pcell{3.3cm}{Ring maps and closed orbit; fixed-step injector with checkpointing} & \pcell{3.1cm}{Symplectic maps, orders 4, 6, 8; sector drift and edges} & \pcell{3.2cm}{Axisymmetric or 3D PIC; cathode image; linac LSC} & \pcell{2.5cm}{Ring, Jacobian, drift, FODO, CLARA} \\
\pcell{1.7cm}{\textsc{Cheetah}~\cite{kaiser2024}} & \pcell{2.7cm}{\textsc{PyTorch}; CPU, CUDA, MPS} & \pcell{3.3cm}{\textsc{PyTorch} AD; differentiable map reduction} & \pcell{3.1cm}{Linear, second-order and drift--kick--drift options} & \pcell{3.2cm}{3D Green-function kick~\cite{dhamrait2026}; no cathode image} & \pcell{2.5cm}{Ring, Jacobian, drift} \\
\pcell{1.7cm}{\textsc{JuTrack}~\cite{wan2025}} & \pcell{2.7cm}{\textsc{Julia}; multithreaded CPU} & \pcell{3.3cm}{\textsc{Enzyme} AD~\cite{moses2020}; first-order Taylor maps} & \pcell{3.1cm}{Symplectic 4th-order 6D tracking} & \pcell{3.2cm}{Transverse spectral model in rectangular pipe} & \pcell{2.5cm}{Ring, Jacobian} \\
\pcell{1.7cm}{\textsc{SciBmad}~\cite{signorelli2026}} & \pcell{2.7cm}{\textsc{Julia}; CPU and CUDA, ROCm, Metal, oneAPI GPUs} & \pcell{3.3cm}{Forward mode (\textsc{ForwardDiff}, GTPSA Taylor series) and reverse mode} & \pcell{3.1cm}{Symplectic integrators, orders 2--8; exact bends} & \pcell{3.2cm}{Planned; not in the release used} & \pcell{2.5cm}{Ring, Jacobian} \\
\pcell{1.7cm}{\textsc{Xsuite}~\cite{iadarola2023}} & \pcell{2.7cm}{Python with compiled kernels; CPU and GPUs} & \pcell{3.3cm}{n.s.} & \pcell{3.1cm}{Symplectic thick and thin maps} & \pcell{3.2cm}{Frozen and PIC models} & \pcell{2.5cm}{Ring, Jacobian} \\
\pcell{1.7cm}{\textsc{ImpactX}~\cite{huebl2022,mitchell2024}} & \pcell{2.7cm}{C++/AMReX and Python; CPU or GPU} & \pcell{3.3cm}{Prototype only~\cite{huebl2026napac}} & \pcell{3.1cm}{$s$-based symplectic maps} & \pcell{3.2cm}{3D rest-frame solve; mesh refinement} & \pcell{2.5cm}{Drift} \\
\pcell{1.7cm}{\textsc{HELIX}~\cite{pathak2026}} & \pcell{2.7cm}{\textsc{PyTorch} AD at fixed energy; C++ PIC, optional GPU FFT} & \pcell{3.3cm}{Reverse-mode Jacobians through PIC cycle} & \pcell{3.1cm}{Linear matrices and multiparticle tracking} & \pcell{3.2cm}{3D FFT PIC and envelope} & \pcell{2.5cm}{FODO} \\
\bottomrule
\end{tabular}
\end{table*}

\subsection{Choosing a model and execution mode}

The comparisons described in this paper distinguish three decisions: which physical approximation is adequate, which precision is required, and which execution mode suits the task at hand. Table~\ref{tab:codes} summarises the model differences relevant here. \textsc{Cheetah} is a close design relative among the codes compared here: both packages use \textsc{PyTorch} and support automatic differentiation. Its reduced linear maps are useful when linear transport is sufficient~\cite{kaiser2024}; the amplitude-dependent ring tunes in this study require nonlinear element models. \textsc{JuTrack} and \textsc{SciBmad} provide alternative differentiable tracking implementations~\cite{wan2025,signorelli2026}. \textsc{JuTrack}'s tune difference in the tested lattice is consistent with the error of its sector-bend map. \textsc{SciBmad} agrees with the reference once its hard-edge fringe fields, which its default method applies, are switched off. Neither result should be interpreted as a general ranking of the codes.

GPU execution becomes economical when many particles share the same lattice calculation. In the ring scaling test, the crossover with single-thread \textsc{Xsuite} occurs at about 50 particles for float32 and 500 for float64. Precision must be chosen depending on the required observable: float32 meets the ring tune tolerance, but not the tighter first-turn map tolerance, and it adds substantial noise to frequency-map diffusion. Double-single arithmetic improves both at extra cost. The measured ratios describe this workstation and the stated resources.

The derivative results make a similar distinction between availability and utility. The ring Jacobian has four outputs, so its reverse-mode cost changes little as more input parameters are added. This benefits 24-parameter optics matching, for example. For the five-parameter injector objective, the gradient costs about six forward evaluations and Bayesian optimisation reaches better objectives in less elapsed time. \textsc{HELIX} reports a different crossover for a PIC calculation~\cite{pathak2026}; however the different outputs, machines and workloads prevent a direct comparison of crossover counts.

The tolerance calculation also suggests a use of derivatives beyond gradient-based search. The linearised budget needs the Jacobian of the observables with respect to every error source, which reverse-mode AD supplies at a cost independent of the number of sources. A tolerance estimate could therefore be updated during an optimisation, instead of by a separate Monte Carlo study once the design is fixed. Each update costs one reverse sweep per observable, against two evaluations per source for finite differences and some tens of tracked machines for a Monte Carlo, which also does not attribute individual sources.

\textsc{Xsuite} and \textsc{ImpactX} provide established alternatives for GPU tracking and collective effects~\cite{iadarola2023,huebl2022}. \textsc{SciBmad} also tracks on GPUs, differentiably, but has no collective effects in the release tested~\cite{signorelli2026}. \textsc{Slipstream} brings supported differentiable paths and field-map injector tracking into the same Python environment as its ring and linac models. \textsc{ASTRA}, \textsc{GPT} and \textsc{elegant} provide independent numerical comparisons. At CLARA, the 14--22~s default-setting runtimes can support repeated scans, while the finer calculations are appropriate for checking selected candidates. The remaining differences in beam size and longitudinal observables must accompany any claim based on emittance agreement alone.

\subsection{Interfaces for digital-twin studies}

The motivating digital-twin architecture separates a machine description, a virtual control system and a simulation module~\cite{brynes2026,king2025}. \textsc{Slipstream} can build an injector from the section-based machine descriptions provided by \textsc{LAURA}, the Lattice Architecture for a Unified Representation of Accelerators~\cite{laura,brynes2026lattice}. It writes openPMD-beamphysics particle files and Twiss tables in the layout used by the simulation module~\cite{huebl2015,simba}. Cavity phase, solenoid field and laser spot size correspond to quantities that can be varied in control-system studies.

These interfaces establish compatibility with that workflow. The present benchmarks measure simulation behaviour and runtime; they do not demonstrate a deployed digital twin, control-system latency or prediction of machine measurements. Scans and fits using these interfaces still require convergence checks and a comparison with the relevant measured observables, as with any code.

\subsection{Model limits and further work}

The remaining discrepancies identify where further work is useful. The one-dimensional, unshielded CSR model leaves differences in the full compressor and in radiation carried between bends. Shielded or three-dimensional models would address regimes beyond the approximation tested here. The quasi-static injector model uses a mean beam energy; an energy-resolved treatment near the cathode would test that approximation more directly. Broader field-map support would allow comparisons with additional machine descriptions.

For derivatives, the demonstrated injector path has fixed emission times and excludes aperture losses. Larger-parameter optimisation studies would test whether its gradient cost becomes economical beyond the five-parameter example. The use of \textsc{Slipstream} in modern optimisation algorithms, such as generative phase space reconstruction~\cite{gpsr}, will also be a useful extension to the existing code, and could provide a simple path for differentiable simulations for collective-effect dominated lattices (with strong CSR or SC signatures).

\section{Conclusions}
\label{sec:conclusions}

\textsc{Slipstream} combines non-linear transport, field-map photoinjector tracking and selected collective-effect models in \textsc{PyTorch}. The extensive benchmarking of \textsc{Slipstream} against standard tools used in the community, across multiple beam physics modalities, shows that it is a suitable tool for accurate simulations, with GPU execution and derivatives that reduce the cost of repeated calculations in the cases tested.
 
The ring benchmarks show close agreement with \textsc{elegant} and other standard tools. GPU execution reduces the cost of tracking large particle ensembles, with the run time per turn scaling well as the number of particles increases, while CPU tracking remains competitive for small workloads. 

Studies on the CLARA injector and bunch compressor show why convergence has to accompany any comparison between codes. While we chose one code in particular (\textsc{ASTRA}) for our injector benchmarking study and \textsc{elegant} / \textsc{Ocelot} for the CSR study, the physics models used across different codes -- particularly in regimes dominated by collective effects -- are not the same. Agreement between codes is only a step towards the ultimate goal: to provide a realistic physical model of a machine at a reasonable speed; nevertheless, the benchmarks are an informative case study that shows how each code behaves as a function of numerical parameters, and that \textsc{Slipstream} is capable of reaching a similar level of accuracy as other codes tested whilst remaining competitive in computational speed. The injector study shows that \textsc{Slipstream} also offers fast, differentiable tracking for in-depth studies in this regime, although at five parameters its gradients did not shorten the optimisation and larger parameter counts remain to be tested, including their use for tolerance estimation during optimisation.
 
The development of \textsc{Slipstream} also provides a documented example of AI-assisted implementation under human direction, assessed through independent numerical comparisons; development took about 12 weeks and roughly 100--150~h of working sessions, with logged Claude usage equivalent to about \$400 at API list prices (Sec.~\ref{sec:ai-development}).

The usage of AI-assisted code developments also has profound implications for the role of simulation codes in this field -- anyone with time and a modest budget can now create an accurate, fast, beam tracking code, with multiple collective effects models. In fact, it is likely that, in the coming months and years the only real differentiation between codes will be their programming language (\textsc{Python}, \textsc{JAX}, \textsc{Julia} etc.) and their choice of accelerator (CPU, GPU, TPU, FPGA, etc.). It may come to pass that \textsc{Slipstream} itself is less relevant than the harness built to validate, benchmark, and ultimately compare it to a suite of its peers.
\begin{acknowledgments}
The authors used the AI tools identified in Sec.~\ref{sec:ai-development} to assist with software development, manuscript drafting and revision, and preparation of plots and illustrations. All use was under human direction and supervision. The authors take responsibility for the software, figures, interpretations and final manuscript. Final manuscript editing was done by the authors.
\end{acknowledgments}

\section*{Data and code availability}

\textsc{Slipstream} is open source under the Apache-2.0 licence. Release v1.0.1, which reproduces the \textsc{Slipstream} results in this paper, is available at \url{https://github.com/astec-stfc/slipstream/releases/tag/v1.0.1}. The benchmark cases, comparison scripts and saved results are in the slipstream-validation repository, \url{https://github.com/astec-stfc/slipstream-validation}.

\bibliography{references}
\end{document}